\documentclass[useAMS,usenatbib]{mn2e}
\usepackage{graphicx}
\usepackage{pslatex}
\usepackage{natbib}
\def\deg{{\rm o}}
\def\idm#1{{\mbox{\scriptsize #1}}}
\newcommand\Chi{{(\chi^2_\nu)^{1/2}}}
\newcommand\stara{{14~Her}}
\newcommand\Ym{\langle Y\rangle}

\def\mean#1{\langle #1\rangle}
\begin{document}
   \title{A dynamical analysis of the 14~Her planetary system}
   \author[K. Go\'zdziewski, C. Migaszewski and M. Konacki]{Krzysztof Go\'zdziewski$^1$\thanks{Toru\'n Centre for Astronomy, Poland}, 
    Cezary Migaszewski$^1$\thanks{Toru\'n Centre for Astronomy, Poland} and
    Maciej Konacki$^2$\thanks{Nicolaus Copernicus Astronomical Center,
    Polish Academy of Sciences, Toru\'n, Poland; Obserwatorium Astronomiczne, 
    Uniwersytet A. Mickiewicza, Pozna\'n, Poland}
   }

   \maketitle
   \begin{abstract}
   Precision radial velocity (RV) measurements of the Sun-like dwarf 
   14~Herculis published by Naef et. al (2004), Butler et. al (2006) and
   Wittenmyer et al (2007) reveal a Jovian planet in a 1760~day orbit and a
   trend indicating the second distant object. On the grounds of dynamical
   considerations,  we test a hypothesis that the trend can be  explained  by
   the presence of an additional giant planet. We derive dynamical limits to the
   orbital parameters of the putative outer Jovian companion in an orbit within
   $\sim 13$~AU. In this case, the mutual interactions between the Jovian
   planets are important for the long-term stability of the system.  The best
   self-consistent and stable Newtonian fit to an edge-on configuration of
   Jovian planets has the outer planet in 9~AU orbit with a moderate
   eccentricity $\sim 0.2$ and confined to a zone spanned by the low-order mean
   motion resonances 5:1 and 6:1. This solution lies in a shallow minimum of
   $\Chi$ and persists over a wide range of the system inclination.  Other
   stable configurations within $1\sigma$ confidence interval of the best fit
   are possible for the semi-major axis  of the outer planet in the range of
   (6,13)~AU and the eccentricity in the range of (0,0.3).  The orbital
   inclination cannot yet be  determined but when it decreases, both planetary
   masses approach  $\sim 10 \mbox{m}_{\idm{J}}$ and for $i \sim 30^{\deg}$ the
   hierarchy of the masses is reversed.  
  \end{abstract}

\section{Introduction}
%
An extrasolar planet around 14~Herculis discovered by the Geneva planet search
team was announced in a conference talk (1998). The star was also monitored by
other planet-hunting teams. The Jovian companion in $\sim 1700$~days orbit was
next confirmed by \cite{Butler2003}. In their new paper,  the discovery team
\citep{Naef2004} published 119 observations revealing a linear RV trend of 
$\sim 3.6$~m/s per year. The single planet Keplerian solution to these
measurements yielded an abnormally large rms of about $14$~m/s. Even if the
drift was accounted for, the rms of the single planet+drift model was $\sim
11.3$~m/s, significantly larger than the mean observational uncertainty
$\mean{\sigma_{\idm{m}}} \sim 7.2$~m/s quoted in that paper. 

Following the hypothesis that the linear trend in the RV data may indicate
other, more distant bodies, in \cite{Gozdziewski2006a}, we merged these
measurements  with  accurate, 35 observations (the mean of $\sigma_{\idm{m}}\sim
3.1$~m/s) by \cite{Butler2003}, spanning the middle part of the combined
observational window. We reanalyzed the full data set of 154 measurements,
covering about $3400$~days, $\sim 2 P_{\idm{b}}$. Using our GAMP method
\citep{Gozdziewski2003e}, i.e., the optimization incorporating  stability
constraints into the fitting algorithm, we searched for stable configurations of
putative two-planet systems which are consistent with the RV data. This search
revealed many stable solutions, in particular two distinct and equally good best
fits with $a_{\idm{c}}\sim 5.8$~AU and $a_{\idm{c}}\sim 9$~AU.  They were
localized in the proximity of low-order mean motion resonances (MMRs), 3:1~MMR
(14~Her$^a$) and 6:1~MMR (14~Her$^b$). Apparently, the companions would be well
separated, but the system would still be active dynamically.  The dynamical maps
computed in the neighborhood of the selected fits revealed a complex structure
of the phase space. We found a clear relation of  chaotic motions to strongly
unstable, self-disrupting  configurations on a short time-scale of $10^4$
periods of the outermost planet.

After the upgrade of HIRES spectrograph and improvements to the data pipeline, 
\cite{Butler2006} have re-analyzed their spectra and published a revised set of
50 observations of 14~Her. Their mean accuracy $\sim 2.4$~m/s is already very
good. Yet, towards the end of the observational window, single data points  have
formal errors as small as 1~m/s. Recently, \cite{Wittenmyer2007}  have also
published 35 additional and independent precision measurements made at the
McDonald Observatory. They have a mean accuracy of 7.5~m/s. The full publicly
available data set is now comprised of 203 observations spanning 4463.2~days and
still does support the presence of a second planetary  object. The single-planet
solution has an rms $\sim 13$~m/s, and  shows a few meter per second excess 
over the mean $\sigma_{\idm{m}} \sim 6$~m/s, added in quadrature to the stellar
jitter $\sim 3.5$~m/s \citep{Butler2006}. Using the Keplerian model of the RV,
\cite{Wittenmyer2007} have found the best-fit solution in the vicinity of the
4:1 mean motion resonance (MMR). However, having in mind the results of our
earlier dynamical analysis \citep{Gozdziewski2006a}, we can suspect that the
kinematic model is not  adequate to properly interpret the RV observations. 

In this paper, we re-analyse the updated high-precision RV data to study and
extend the kinematic model by \cite{Wittenmyer2007}. We also revise and
correct the conclusions from our previous paper \citep{Gozdziewski2006a} that
were derived on the basis of a smaller and less precise RV data set.  The goal
of our work is to derive dynamical characteristics of the putative planetary
system and to place limits on its orbital parameters.   Assuming a putative
configuration close to the 4:1~MMR, the outermost  orbital  period would be
$\sim 18$--$20$~yr. Hence, the currently available  data would cover only a part
of this period and a full characterization of  the orbits would still require
many years of observations. One cannot expect  that in such a case the orbits
can be well determined. Nevertheless, we show  in this work, that the modeling
of the RV data when merged with a dynamical analysis and mapping of the phase
space of initial conditions with fast-indicators enable us to derive meaningful
limits to the orbital parameters. Such an extensive dynamical study of this
systems has not yet been done.

In order to take into account the RV variability  induced by the stellar
activity, the internal errors of the RV measurements are rescaled by the
jitter   added in quadrature: 
$
 \sigma^2 = \sigma_{\idm{m}}^2 + \sigma_{\idm{j}}^2,
$ 
where $\sigma_{\idm{m}}$ and $\sigma_{\idm{j}}$ are the mean of  the internal
errors and the adopted estimate of the stellar jitter.  The
14~Herculis is a quiet star with $\log R'_{\idm{HK}}=-5.07$ \citep{Naef2004}
thus it is reasonable to keep the safe estimate of  $\sigma{\idm{j}}=3.5$~m/s
\citep{Butler2006}. The mass of the parent star varies by $\sim 10\%$ in
different publications. In this work, we adopt the canonical mass
$1\mbox{M}_{\sun}$, although we also did some calculations for the  mass of
$0.9\mbox{M}_{\sun}$ and $1.1\mbox{M}_{\sun}$.

\begin{figure*}
   \centerline{
   \vbox{
   \hbox{
       \hbox{\includegraphics[width=6.0in]{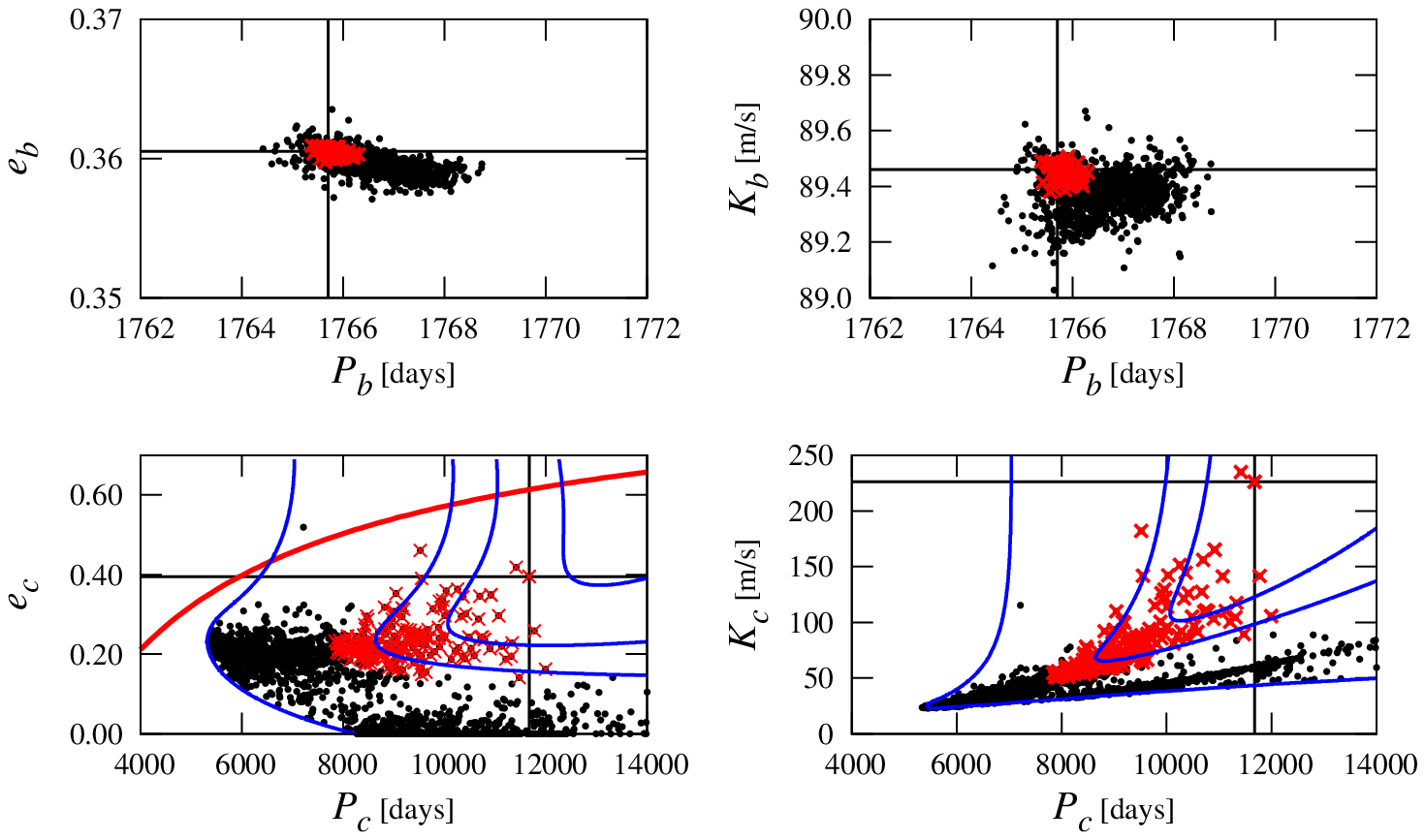}}
   }
   \hbox{
       \hspace*{1.2cm}
       \hbox{\includegraphics[width=2.56in]{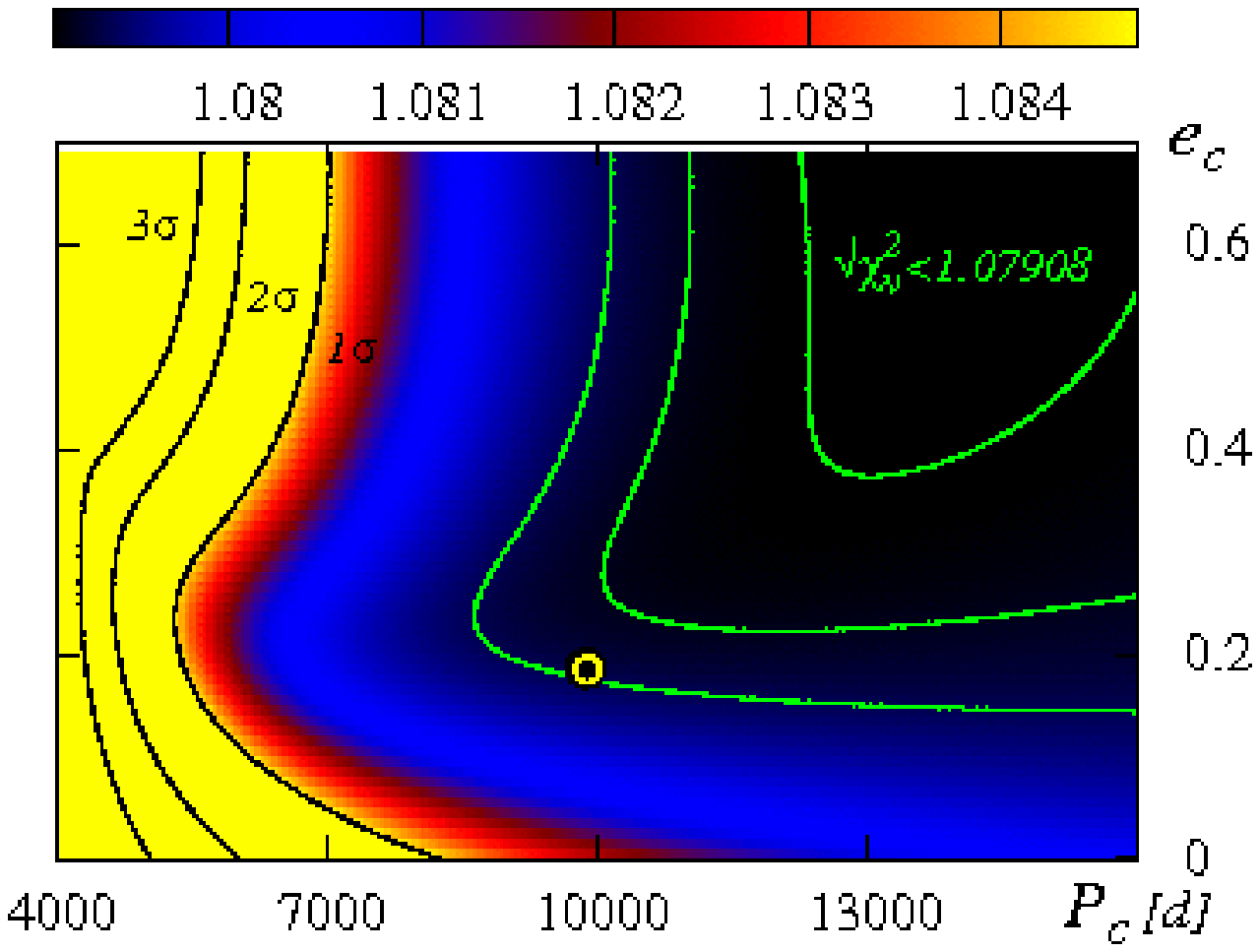}}
       \hspace*{0.8cm}
       \hbox{\includegraphics[width=2.56in]{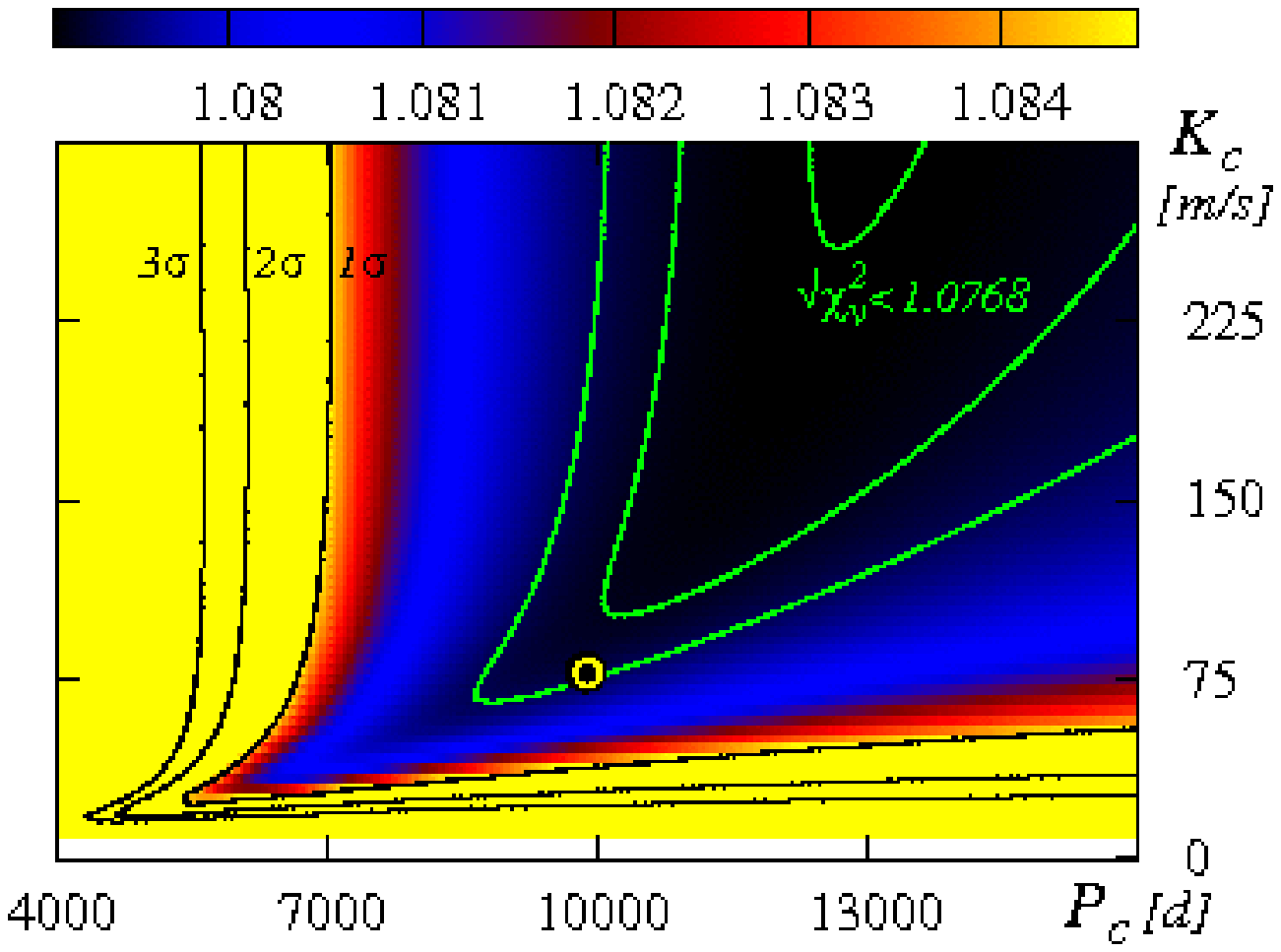}}
   }
   }
   }
   \caption{
   The best fit solutions to the two-planet model of the RV data of \stara{} in
   the ($P,e$)- and ($P,K$)-plane. In the search,  a distant companion to the
   known Jovian planet in $\sim 1760$~days orbit is assumed. Upper plots are for
   the inner planet, middle plots are for the outer one.  The best-fit found
   with the hybrid code in the range of moderate orbital periods of the outer
   planet, with $\Chi \sim 1.085$, is marked with intersecting lines. Its
   elements, in terms of the parameter tuples of Eq.1, $(K\mbox{[m/s]},
   P\mbox{[days]}, e,  \omega\mbox{[deg]},T_{\idm{p}}-T_0\mbox{[days]})$,   are
   (89.460, 1765.743, 0.360,  22.108, 1375.03) for the inner and (226.2263,
   11675.333, 0.395, 182.006, 6115.352 ) for the outer planet, respectively,
   with the velocity offsets 
   ($V_{\idm{N2004}},V_{\idm{B2006}},V_{\idm{W2007}}$)   $=$ 
   (-129.105, -132.345, -108.384)~m/s for the data in \citep{Naef2004}, \citep{Butler2006} and
   \citep{Wittenmyer2007}, respectively;  the epoch $T_0=$~JD2,450,000.  
   Dots are for $1\sigma$  solutions with $\Chi \in [1.0848,1.090)$ and an
   rms $\sim 8.17$~m/s,
   red crosses are for the the best fit solutions with $\Chi \sim 1.0848$. 
   The red curve in ($P_{\idm{c}},e_{\idm{c}}$)-plane marks an approximate
   collision line of the orbits determined from $a_{\idm{b}} (1+e_{\idm{b}}) =
   a_{\idm{c}} (1-e_{\idm{c}})$, with $a_{\idm{b}},e_{\idm{b}}$ fixed at their
   best-fit values. 
   Bottom panels are for the scans of $\Chi$ at the selected
   parameter planes. Contours (also shown in the upper panels,
  for a reference) mark the limits of
   $\Chi$ corresponding to $1\sigma,2\sigma$ and $3\sigma$ of the
   best fit solution.
   Circle mark the position of the best $N$-body solution (see Sect.~4).
   See the text for remaining details.
   }
\label{fig:fig1}
\end{figure*}
\section{Modeling the RV data -- an overview}
%
The standard formulae by~\cite{Smart1949} are  commonly used to model the RV
signal. Each planet in the system contributes to the reflex motion of the star
at time $t$ with:   \begin{equation} V_{\idm{r}}(t) = K [ \cos (\omega+\nu(t)) 
+ e \cos \omega] + V_0, \label{eq:eq1}   \end{equation} where $K$ is the
semi-amplitude, $\omega$ is the argument of pericenter, $\nu(t)$ is the true
anomaly (involving implicit dependence on the orbital period $P$ and time of
periastron passage $T_{\idm{p}}$), $e$ is the eccentricity, $V_0$ is the
velocity offset. We interpret the primary model parameters
$(K,P,e,\omega,T_{\idm{p}})$ in terms of the Keplerian elements and minimal
masses related to Jacobi coordinates \citep{Lee2003,Gozdziewski2003e}.  

In our previous works, we tested and optimized different tools used to explore
the multi-parameter space of $\chi^2$. In the case when $\chi^2$ has many local
extrema, we found that the hybrid optimization  provides particularly  good
results \citep{Gozdziewski2004, Gozdziewski2006b}. A run of our code starts the
genetic algorithm \citep[GA,][]{Charbonneau1995} that basically  has a global
nature and requires only the $\chi^2$ function and rough parameter ranges. GA
easily permits for a constrained optimization within prescribed parameter bounds
or for a penalty term to $\chi^2$ \citep{Gozdziewski2006a}.  Because the best
fits found with GAs are not very accurate in terms of $\chi^2$, we refine  a
number of ``fittest individuals'' in the final ``population'' with a relatively
fast local method like the simplex of Melder and Nead~\citep{Press1992}.   The
use of simplex algorithm is a matter of choice and in principle we could use 
other fast local methods. However, a code using non-gradient methods works  with
the minimal requirements for the user-supplied information, i.e., the model
function, usually equal to $\chi^2$ [its inverse  $1/\chi^2$ is the {\em fitness
function} required by GAs].   The  repeated runs  of the hybrid code provide an
ensemble of the best-fits that helps us to detect local minima of $\chi^2$, even
if they are distant in the parameter space. We can also obtain reliable
approximation to the errors of the best-fit parameters \citep{Bevington2003}
within the $1\sigma$, $2\sigma$ and $3\sigma$  confidence levels of $\chi^2$ at
selected  2-dim parameter planes,  as corresponding to appropriate increments of
$\chi^2$. The hybrid approach will be called the algorithm~I from hereafter.

However, in this paper we deal with a problem of only a partial coverage of the
longest orbital period by the data. In such a case $\chi^2$ does not have a  
well defined minimum or the space of acceptable solutions is  very ``flat'' so 
the confidence levels may cover large ranges of the parameters.  In such a case,
to illustrate {\em the shape} of $\chi^2$ in the selected 2-dimensional
parameter planes, we use a complementary approach that relies on systematic
scanning of the parameter space with the fast Levenberg-Marquardt (L-M)
algorithm \citep{Press1992}. Usually, as the parameter plane for representing
such scans, we choose the semimajor-axis---eccentricity $(a,e)$ plane of the
outermost planet.  We fix $(a,e)$ and then search for the best fit, with the
initial conditions selected randomly (but within reasonably wide parameter
bounds).  Then the L-M scheme ensures a rapid convergence. This approach is
called the algorithm~II from hereafter. In fact, we already used it in
\cite{Gozdziewski2003e}. The method is  CPU-time consuming and may be
effectively applied in low-dimensional problems but in reward it can provide a
clear and global picture of the parameter space.

Moreover, the RV measurements can be polluted by many sources of error, like 
complex systematic instrumental effects, short time-series of the observations,
irregular sampling due to observing conditions, and stellar noise.  The problem
is even more complex when we deal with models of resonant or strongly
interacting planetary systems. It is well known that the kinematic model of the
RV is not adequate to describe the  observations in such instances. Instead, the
self-consistent $N$-body Newtonian model should be applied
\citep{Rivera2001,Laughlin2001}. (Note that the hybrid optimization can be used
to explore the parameter space for both models).   Still, the best-fit solutions
may be related (and often do) to unrealistic quickly disrupting configurations
\citep{FerrazMello2005,Lee2006,Gozdziewski2006a}. In such a case one should
explore the dynamical stability of the planetary system in a neighborhood of the
best-fit configuration. One can do that with the help of direct numerical
integrations or by resolving the dynamical character of orbits in terms of the
maximal Lyapunov exponent or the diffusion of fundamental frequencies. Hence,
the most general way of modeling the RV data should involve an elimination of
unstable (for instance, strongly chaotic) solutions during the fitting
procedure. We described  such an approach (called GAMP) in our previous works. 
In particular, we made attempts to analyze the old data set of 14~Her as well as
other stars hosting multiple planet systems \citep{Gozdziewski2006a}.

\section{Keplerian fits}
%
In order to compare our fits with the results of \cite{Wittenmyer2007}, we
carried out the hybrid search for the two-planet edge-on configuration.    In
the experiment, we limited the orbital periods to the range of 
$[1000,14000]$~days and the eccentricities to the range of $[0,0.9]$.  The
choice of the upper limit of orbital periods followed the conclusions of
\cite{Wittenmyer2007}. According to the results of adaptive-optics imaging by
\cite{Luhman2002} and \cite{Patience2002}, the parent star has no stellar
companion beyond $9$ --- $12$~AU, so we try to verify the hypothesis about a
short-period, low-mass companion \stara{}~c up to that distance. The RV offsets
are different for each of the three instruments and are included as free
parameters in the model together with the tuples of $(K,P,e,\omega,T_{\idm{p}})$
for each companion. Note that we calculate the RV offset  $V_{\idm{N2004}}$ 
with respect to the simple mean of all RV measurements from the set published by
\cite{Naef2004}. 

The results of the hybrid search (algorithm~I) are illustrated in upper panels
Fig.~\ref{fig:fig1} (note that before plotting,  the fits are sorted with
respect to the smaller semi-major axis).  The best-fit found with this method
has $\Chi\simeq 1.085$ and an rms $\simeq 8.17$~m/s. Its parameters are given in
the caption to Fig.~\ref{fig:fig1}.  Clearly, there is no isolated best-fit
solution but rather a huge number of acceptable fits. The  ratio of the orbital
periods within  $1\sigma$ confidence interval of the best fit may be as low as
$\sim 3$. The orbital period in the best fit is $\sim 12000$~days and
$P_{\idm{c}}/P_{\idm{b}}\sim 7$. This is roughly consistent with the work of
\cite{Wittenmyer2007} who quote the best-fit configuration close to the 4:1~MMR
and with a similar rms $8.17$~m/s. Yet  the fits are not constrained with
respect to $P_{\idm{c}}$. Many values between $[5000,14000]$~days are equally
good in terms of $1\sigma$ confidence interval of the best-fit solution. 
Moreover, it means that the system may be confined to a zone spanned by many
low-order mean motion resonances of varying width in the
$(P_{\idm{c}},e_{\idm{c}})$-plane. In such cases, the  mutual interactions
between the planets are important for the long-term stability.

In order to illustrate the shape of $\Chi$ in more detail, we also applied the
algorithm~II for the systematic  scanning of that function in selected parameter
planes (see lower panels in Fig.~\ref{fig:fig1}). This has lead to even better 
statistics of the best fits and a very clear illustration of the
$\Chi$-surface. The panels  reveal that $\Chi$ of 2-Kepler model has no minimum
in the examined ranges of the outermost periods, $P_{\idm{c}} < 14,000$~days.

Along a wide valley, the best fits with an increasing $P_{\idm{c}}$ have also
rapidly increasing $e_{\idm{c}}$. For $P_{\idm{c}} \sim 14,000$~days, they reach
the collision zone of the orbits.  In this strongly chaotic region a number of
MMRs of the type $p:1$ with $p>7$ overlap. It justifies  a-posteriori our choice
of the search limit $P_{\idm{c}} \sim 14,000$~days. Note, that the two searches
are in an excellent accord that is illustrated with the $\Chi$ smooth contour
levels overlayed on the results of the hybrid search. In that case, the scanning
of parameter space helps to understand the shape of $\Chi$ in detail.

The $1\sigma$ scatter of solutions around the best-fit (Fig.~\ref{fig:fig1})
provides realistic error estimates. Although the parameters of the inner planet
are very well fixed, the elements of the outer Jovian planet have much larger
formal errors. The kinematic fits do not provide an upper bound to the
semi-major axis of the outer planet. Both methods reveal exactly the same
parameters of the inner companion (see Fig.~\ref{fig:fig1}). Obviously, certain
discrepancy between the outcomes of algorithm~I and algorithm~II is related to a
``flat'' shape of $\Chi$ in the $(a_{\idm{c}},e_{\idm{c}})$-plane and to a worse
efficiency of simplex in detecting shallow minima compared to the gradient-like
L-M algorithm.  When $\Chi$ does not have a well localized minimum, it is very
difficult to resolve its behaviour with any non-gradient method.

\section{Self-consistent Newtonian fits}
\begin{figure*}
   \centerline{
   \hbox{
       \hbox{\includegraphics[width=2.56in]{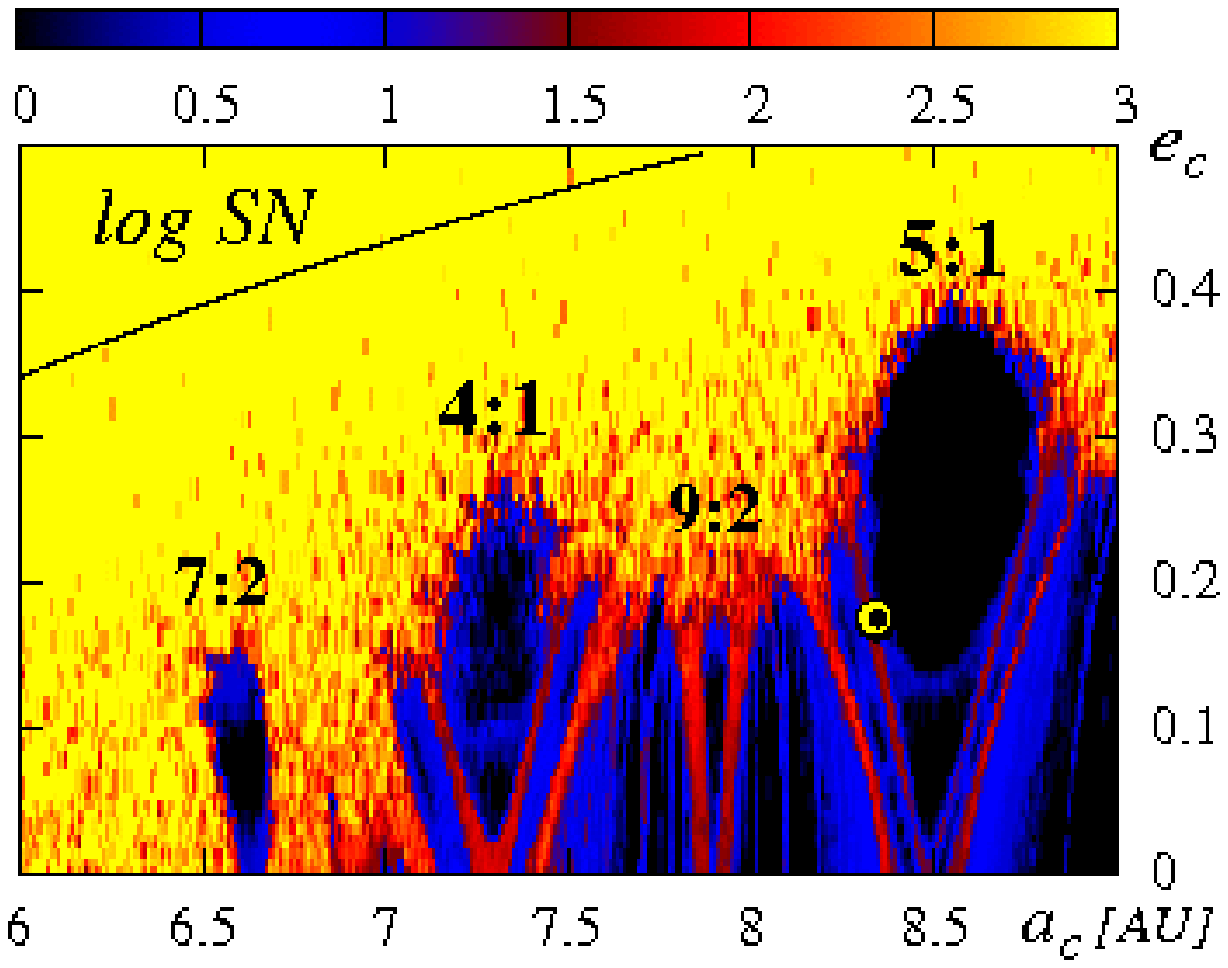}}
       \hspace*{1cm}
       \hbox{\includegraphics[width=2.56in]{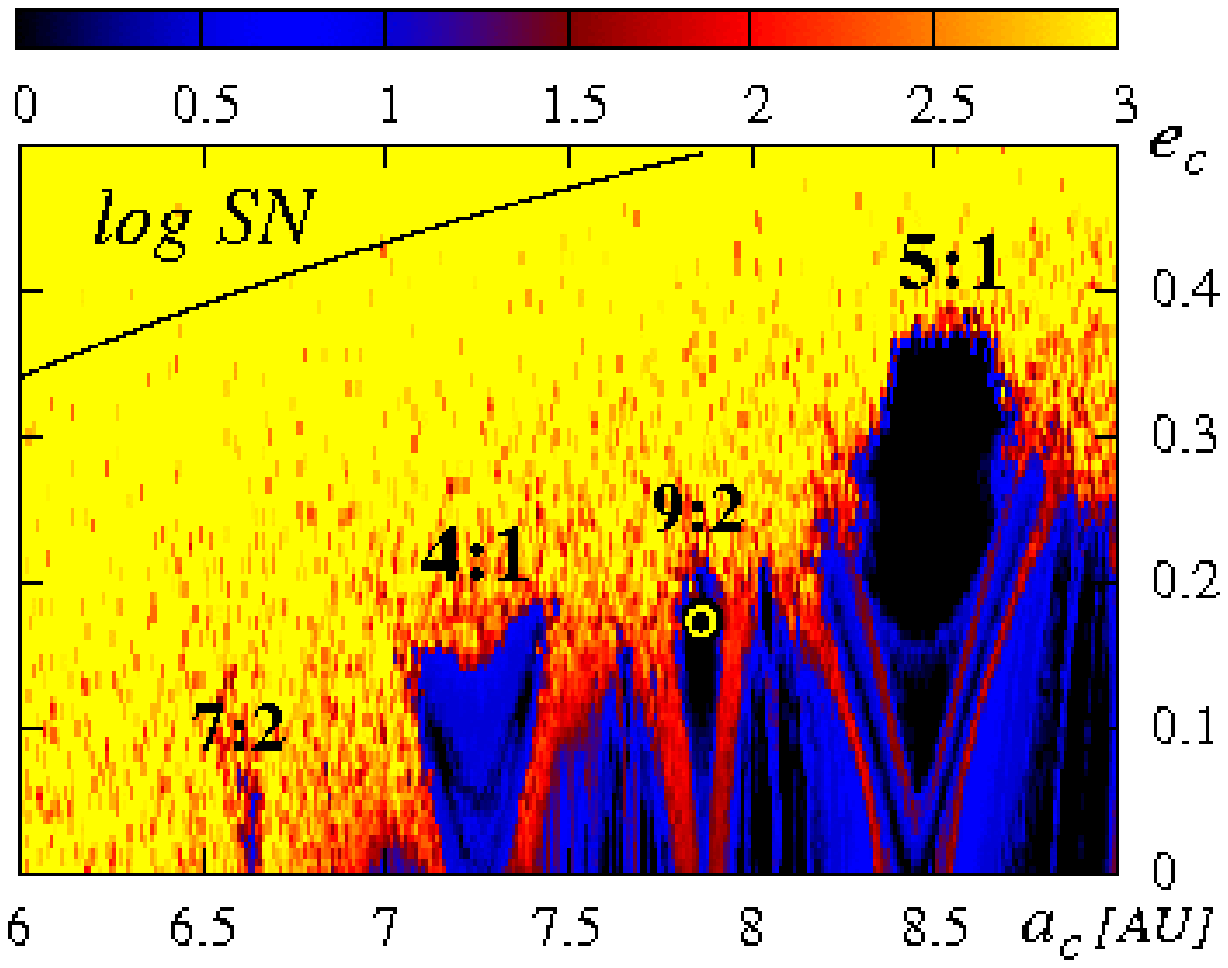}}
   }
   }
   \caption{
   The dynamical maps in terms of the Spectral Number ($\log SN$) in the
$(a_{\idm{c}},e_{\idm{c}})$-plane for the two-planet Newtonian models of the
\stara{} system.  Colors used in the $\log SN$ map classify the orbits --- black
indicates  quasi-periodic regular configurations while yellow strongly chaotic
ones.  The maps are computed for the following osculating elements of the Jovian
planets at the epoch of the first observation $t_0$, in terms of
tuples                                                  
$(m~[m_{\idm{J}}], a~[\mbox{AU}], e, \omega~[\mbox{deg}], {\cal M}(t_0)~[\mbox{deg}])$:
the left panel is for
 (4.975,    2.864,    0.359,   22.281,  330.365) of the inner
 planet, and
 (5.618,    8.343,    0.174,  188.543,   72.091) of the outer planet,
 respectively;
the right panel is for
 (4.974,    2.863,    0.359,   22.310,  330.314) of the inner planet, and
 (4.558,    7.863,    0.173,  187.674,   64.417) of the outer planet,
respectively. The large circles mark the  parameters of the best
fits.   The thin line marks the collision curve for planets b and~c,   as
determined by $a_{\idm{b}} (1+e_{\idm{b}}) = a_{\idm{c}} (1-e_{\idm{c}})$. The
low-order MMRs of the planets b and c are labeled.  The integrations are conducted
over $\sim 10^4 P_{\idm{c}}$. The resolution is $500\times120$ data points.
}
\label{fig:fig2}%
\end{figure*}

\begin{figure*}
   \centerline{
   \vbox{
   \hbox{
       \hbox{\includegraphics[width=2.7in]{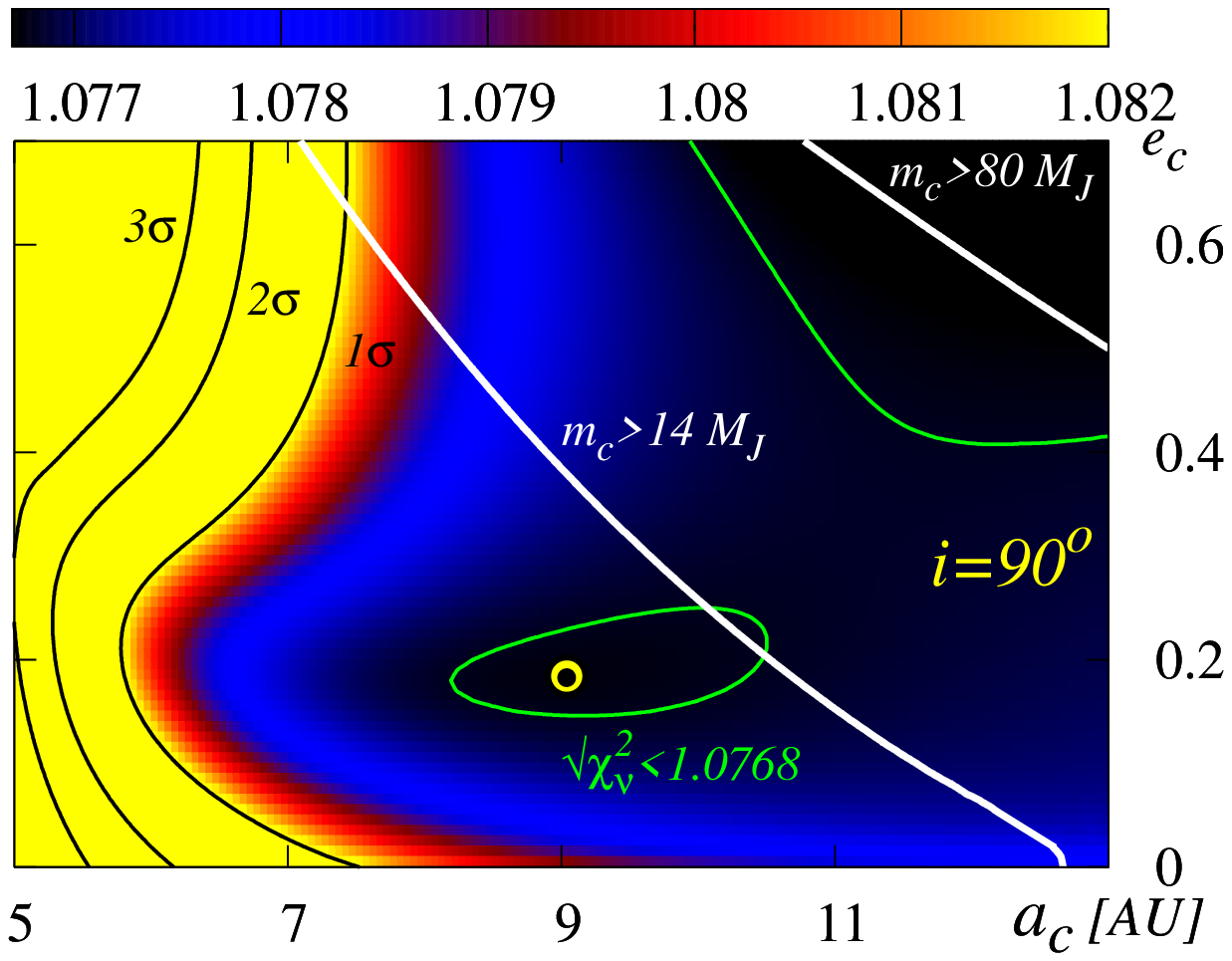}}      
       \hspace*{0.4cm}     
       \hbox{\includegraphics[width=2.65in]{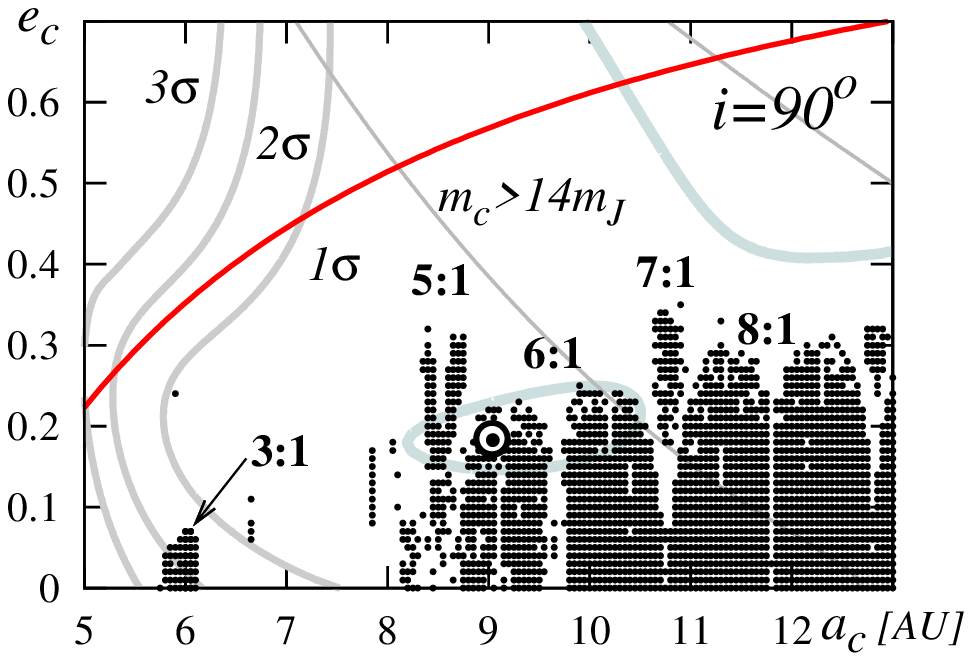}}
   }
   \hbox{
       \hbox{\includegraphics[width=2.7in]{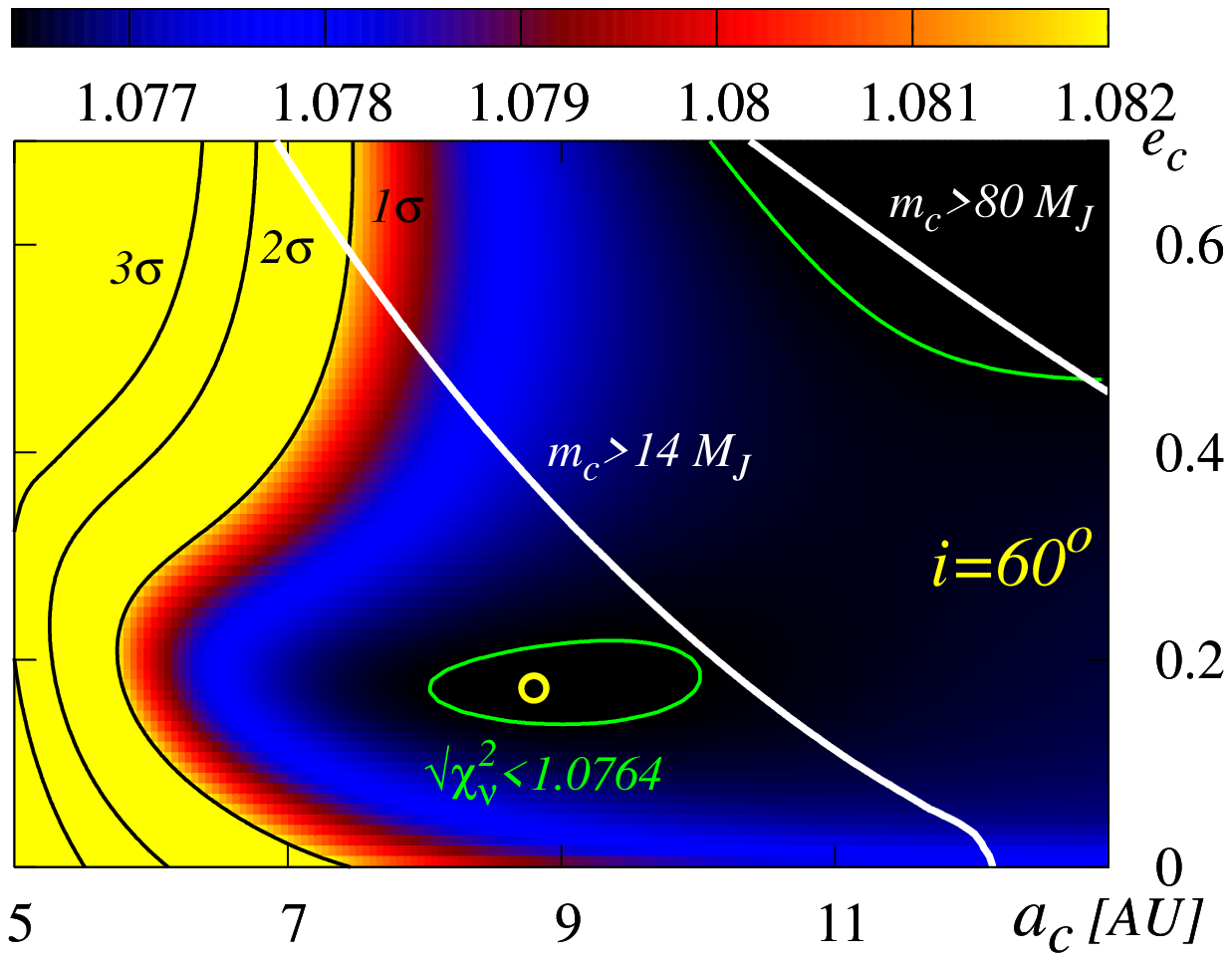}}      
       \hspace*{0.4cm}   
       \hbox{\includegraphics[width=2.65in]{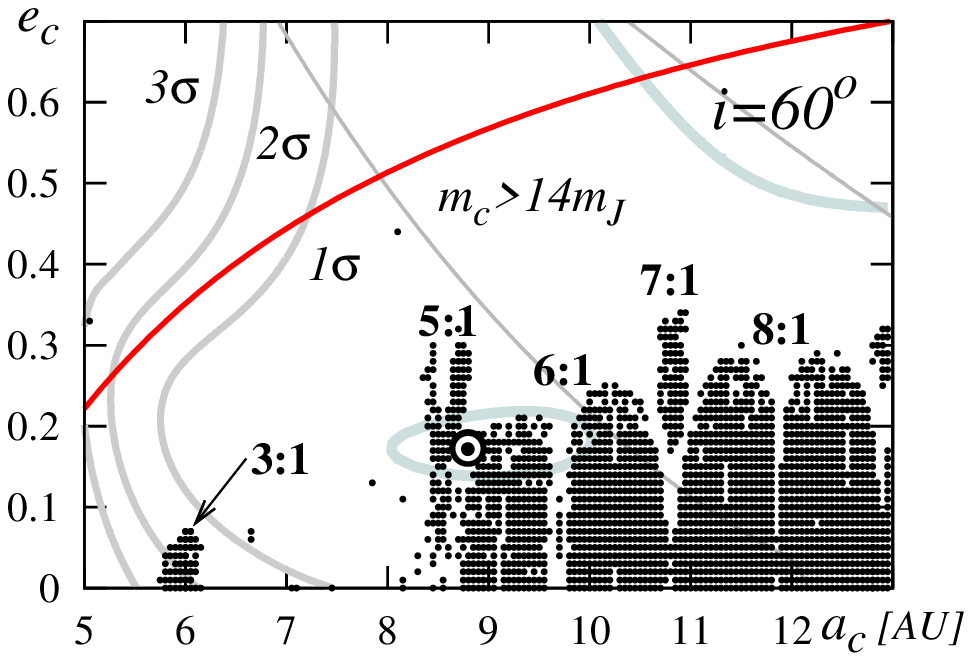}}
   }
   \hbox{
       \hbox{\includegraphics[width=2.7in]{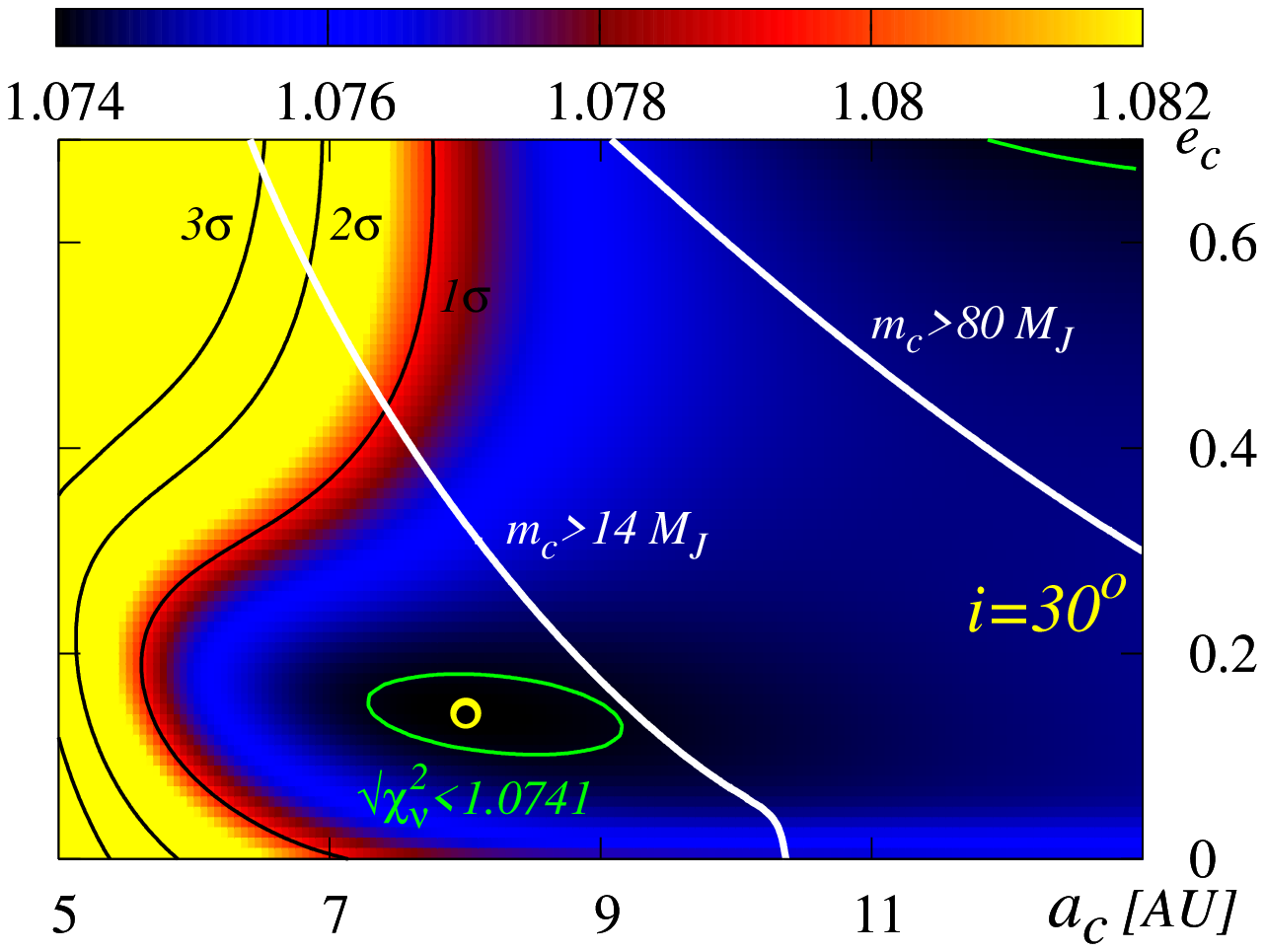}}      
       \hspace*{0.4cm}
       \hbox{\includegraphics[width=2.65in]{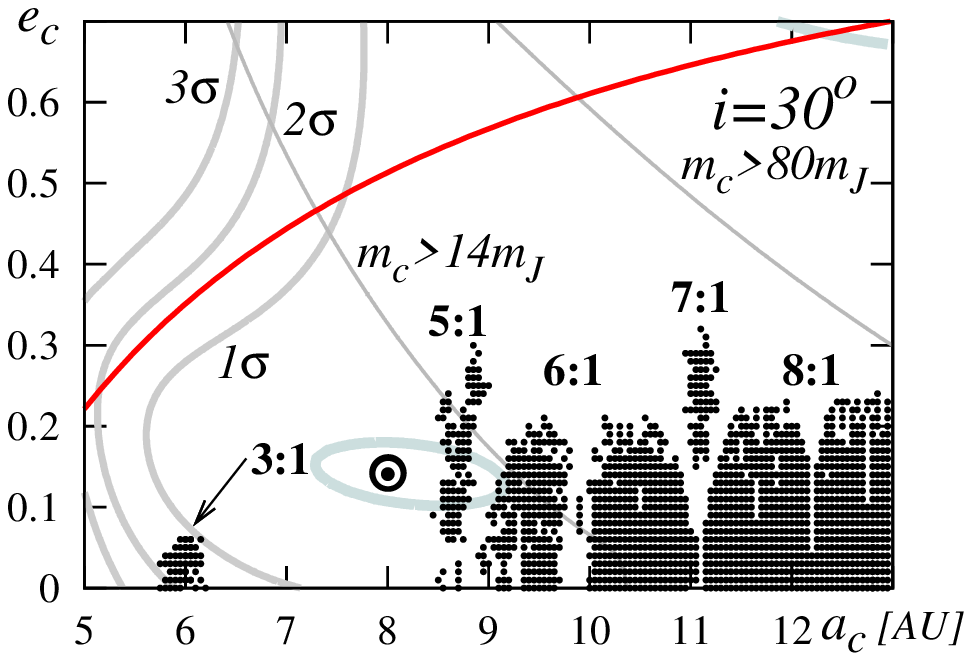}}
   }
   }
   }
   \caption{The panels in the left column are for the Newtonian best-fit model
   with varying fixed inclination of the coplanar system $i$ and illustrated as
   projections onto the $(a{\idm{c}},e_{\idm{c}})$-plane of initial osculating
   elements at the epoch of the first observation.  In each panel, the best fit
   is marked with a circle. The quality of the fits within formal  $1\sigma,
   2\sigma$ and $3\sigma$ confidence interval of the best fit, in terms of their
   $\Chi$ is marked with contours and labeled in subsequent panels.  White thick
   lines corresponds to mass levels of a brown-dwarf and sub-stellar companion,
   respectively. These solutions are derived with algorithm~II. The panels in
   the right column are for the stability analysis of the best-fit
   configurations. See the text for more explanation.
   }
\label{fig:fig3}
\end{figure*}

\begin{figure*}
   \centerline{
   \vbox{
   \hbox{
       \hbox{\includegraphics[width=2.56in]{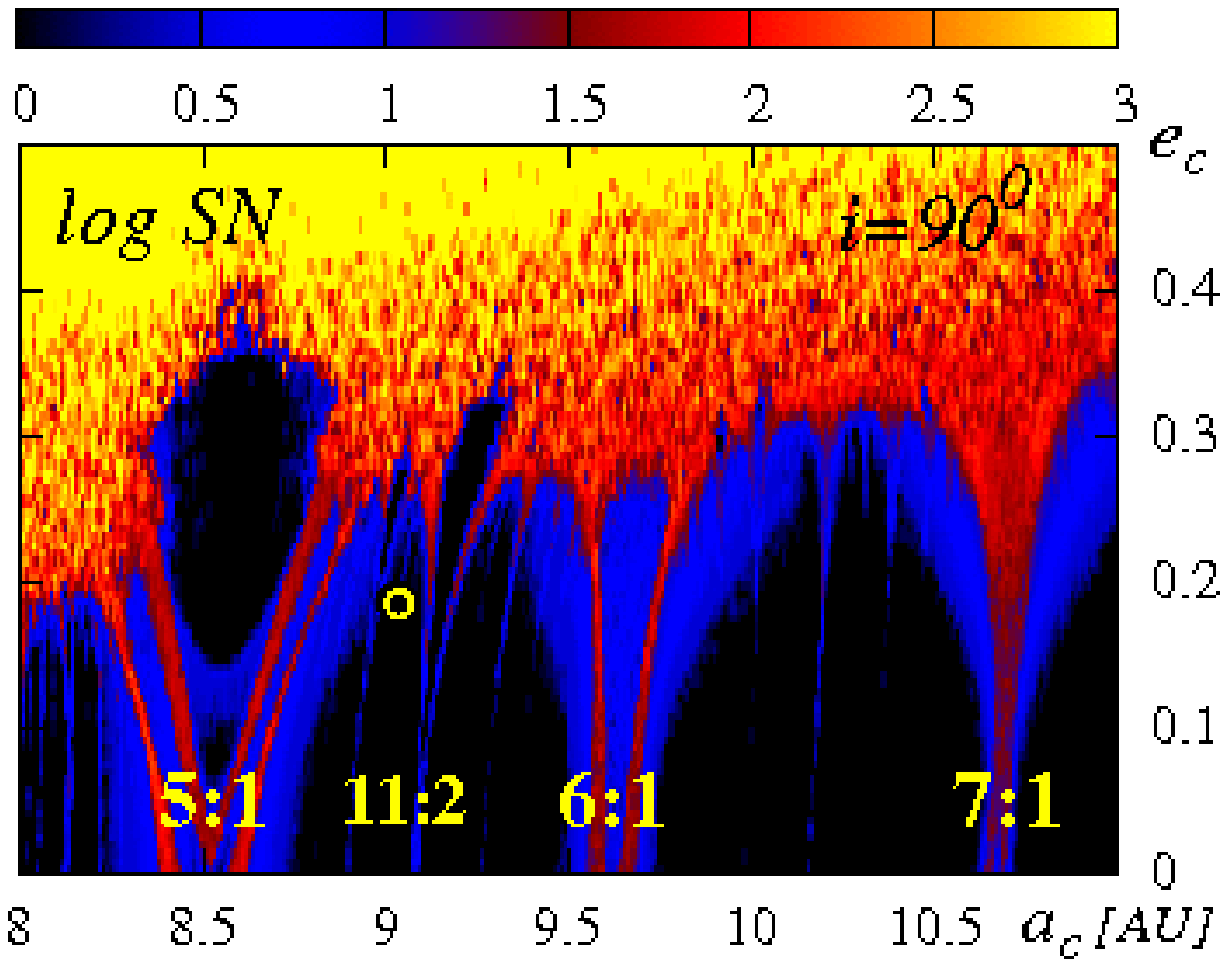}}
       \hspace*{1cm}
       \hbox{\includegraphics[width=2.56in]{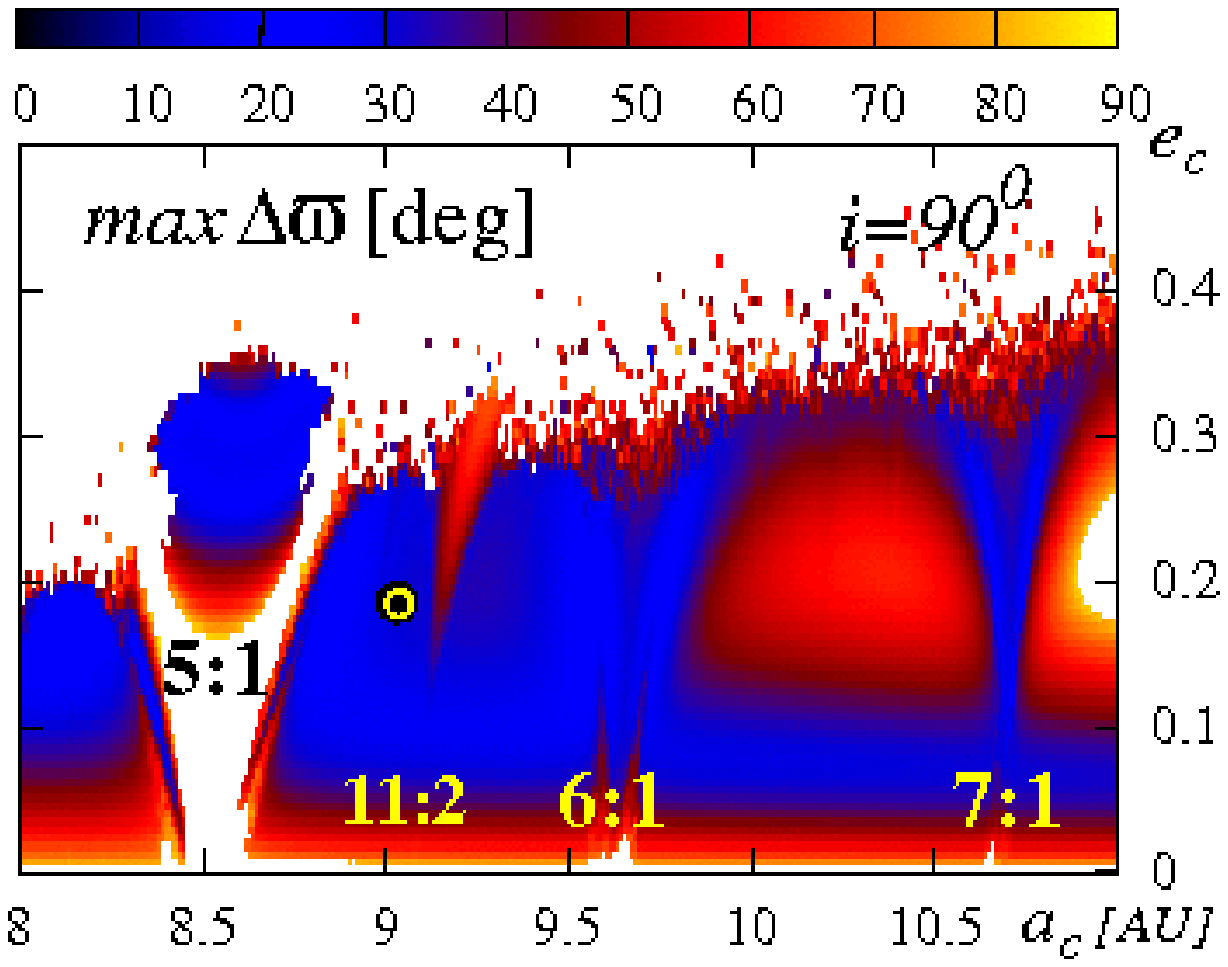}}
   }
   \hbox{
       \hbox{\includegraphics[width=2.56in]{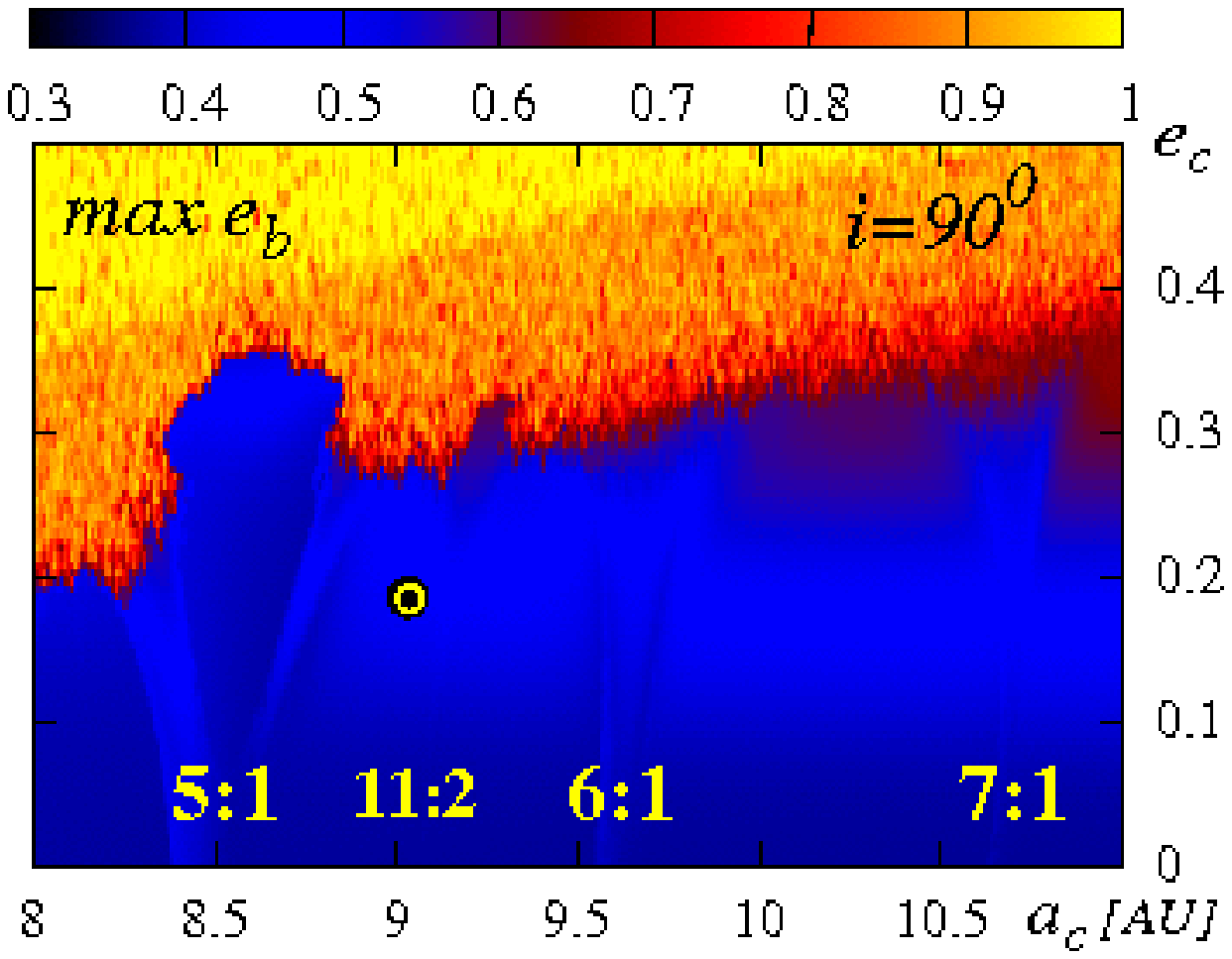}}
       \hspace*{1cm}
       \hbox{\includegraphics[width=2.56in]{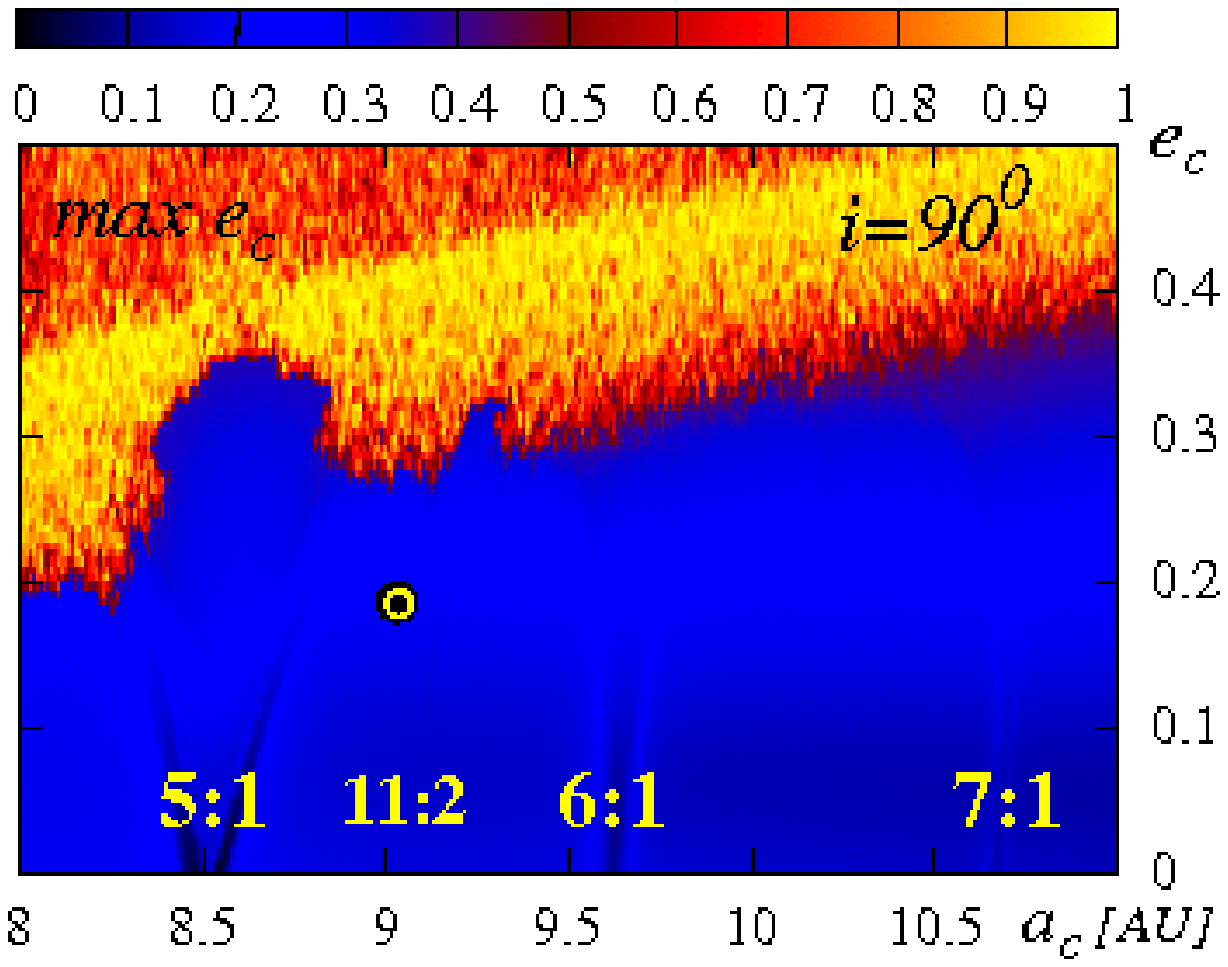}}
   }
   }
   }
   \caption{
The dynamical maps in the $(a_{\idm{c}},e_{\idm{c}})$-plane in terms of the
Spectral Number ($\log SN$, the top-left panel),  $\max \Delta\varpi$  relative
to the libration mode $180^{\circ}$ (the top-right panel) and  $\max
e_{\idm{b,c}}$ (the bottom panels) for the two-planet edge-on coplanar Newtonian
best-fit model of the \stara{} system (Table~1).  Colors used in the $\log SN$
map classify the orbits --- black indicates   quasi-periodic regular
configurations while yellow strongly chaotic ones.  The large crossed circles
mark the parameters of the fit.  The low-order MMRs of the planets b and c are
labeled.  The integrations are conducted over  $\sim 10^4 P_{\idm{c}}$.  The
resolution is $500\times120$ data points. 
}
\label{fig:fig4}%
\end{figure*}

%
In order to compare the outcome of kinematic modeling with the self-consistent
$N$-body one, at first, we transformed the ensemble of Keplerian fits to
astro-centric osculating elements \citep{Lee2003,Gozdziewski2003e}. These fits
became the initial conditions for the L-M algorithm employing the $N$-body model
of the RV. In this experiment, the mass of the star was $M_{\star} =
1~M_{\sun}$. Curiously, we found that the procedure often converged to the two
best-fits, both having similar $\Chi \sim 1.083$, an rms $\sim 8.15$~m/s, and
$a_{\idm{c}}\sim 7.8$~AU and $a_{\idm{c}}\sim 8.4$~AU, respectively.  To shed
some light on the orbital stability of these solutions, we computed the
dynamical maps in the neighborhood of the best-fits in terms of the Spectral
Number \citep[$\log SN$,][]{Michtchenko2001}. We chose $a_{\idm{c}}$ and
$e_{\idm{c}}$ as the map coordinates, keeping other orbital parameters at their
nominal values. The results are shown in Fig.~\ref{fig:fig2}. The left panel is
for the unstable fit located in the separatrix of the 5:1~MMR. The right panel
is for the stable solution in the center of 9:2~MMR.  These dynamical maps prove
that the mutual interactions between planets are  significant --- even small
changes of initial parameters modify the shapes of low-order MMRs which overlap
already for $e_{\idm{c}} \sim 0.2-0.3$. Besides, the chaotic motions are related
to strongly unstable configurations (see also Fig.~\ref{fig:fig3},
\ref{fig:fig4}).

As in the case of the Keplerian model, to resolve  the topology of $\Chi$ in
detail, we performed similar  searches for the best-fit configurations  using
the $N$-body model of the RV.  Again, the results of the hybrid algorithm are
consistent with the algorithm~II. However, to conserve space, we only show the
$\Chi$-maps obtained by scanning the ($a_{\idm{c}},e_{\idm{c}}$)-plane. The
results are illustrated in the left column panels of Fig.~\ref{fig:fig3}.

These scans are computed for three different and fixed values of  the
inclination of the putative orbital plane of the system, $i=90^{\circ}$,
$60^{\circ}$ and $30^{\circ}$, respectively. In the maps, we mark a few contour
levels of $\Chi$ that correspond to $1\sigma, 2\sigma$ and $3\sigma$-levels of
the best fit solution denoted with a circle in every panel.

Overall, the topology of $\Chi$ does change, nevertheless qualitatively certain 
features are common. Contrary to the Keplerian case,  we can detect a shallow
minimum of $\Chi$ close to (9~AU,0.2) and a hint of another minimum localized
far over the collision line. The white thick lines on the panels mark mass
limits of $m_{\idm{c}}$.  Clearly, for $a_{\idm{c}}>9$~AU and for moderate
eccentricity, $m_{\idm{c}}$ approaches the brown-dwarf mass limit and substellar
mass  80~$m_{\idm{J}}$ for $\sim 12$~AU.   Thus a brown-dwarf could exist in the
system beyond $10$-$11$~AU depending on the inclination and eccentricity ---
and  the fits preclude a low-mass planetary object over the 14~$m_{\idm{J}}$
level. 

\begin{figure}
   \hbox{\includegraphics[width=3.3in]{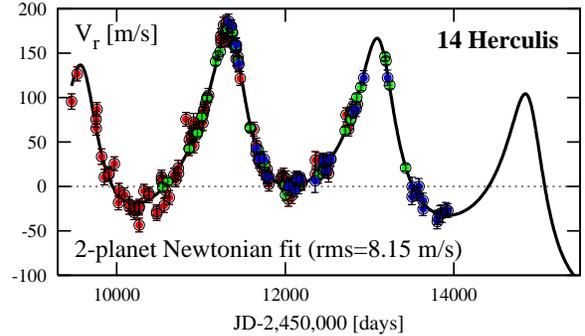}}
   \caption{
   Synthetic RV curve of the self-consistent best fit to the two-planet model.
   The osculating elements at the epoch of the first observation are given in
   Table~1.  This fit yields $\Chi \sim 1.082$ and an rms $\sim 8.15$~m/s.
   Observations from Naef et al. (2004),  Butler et al. (2006) and Wittenmyer et
   al. (2007) are marked  with  red, blue and green circles, accordingly.
   }
\label{fig:fig5}
\end{figure}

Note, that the best-fit for the nominal edge-on configuration 
with $i=90^{\circ}$ turns out to be {\em unstable} but its parameters are very
close to {\em a stable} solution given in Table~1; its synthetic curve is
illustrated in Fig.~\ref{fig:fig5}.

To characterize the stability of the best-fit solutions, we computed their MEGNO
fast indicator \citep{Cincotta2000}, $\Ym$, (i.e., an approximation of the
maximal Lyapunov exponent) up to $\sim 2\times 10^5~P_{\idm{c}}$ using a fast
and accurate symplectic algorithm \citep{Gozdziewski2005b}. The total
integration time $3$-$4$~Myr is long enough to detect even weak unstable MMRs.

The best-fits that passed the MEGNO test (i.e. that are regular over the
integration time) are marked with black dots in the right panels of
Fig.~\ref{fig:fig3}.  For a reference, the $\sigma$-levels of $\Chi$ from the
respective left-panels are marked in these plots as well. The red curves are for
the collision line of the orbits computed for well constrained and fixed
parameters of the inner planet. Overall, the stable solutions reveal the
structure of the MMRs that we have already seen in the dynamical maps computed
for fixed orbital parameters  (see Fig.~\ref{fig:fig2}). The semi-major axis of
the outer planet cannot be yet constrained well and may vary between $6$~AU and
(at least) 13~AU.  Moreover there is a rather evident although irregular border
$e_{\idm{c}} \sim 0.3$ of all solutions, up to $3\sigma$.  Besides, we found
that in the relevant range of $a_{\idm{c}}$, the mean longitude
$\lambda_{\idm{c}} = \varpi_{\idm{c}} + {\cal M}_{\idm{c}} \sim
(200^{\circ},300^{\circ})$ within the $1\sigma$ interval of the best fit,
increasing along quasi-parabolic curve as the function of $a_{\idm{c}}$.
Simultaneously, $\lambda_{\idm{b}} \sim 352.6^{\circ} \pm 1.5^{\circ}$. From the
dynamical point of view, this provides significant limits on the relative mean
phases of the planets.

The results for two-planet 14~Her system in \cite{Gozdziewski2006a} are
basically in accord with the present work although not all conclusions are
confirmed. The two isolated minima of $\Chi$ found in that paper can be
identified in this work too.  In particular it concerns the narrow 3:1~MMR
island. The second solution close to the 6:1~MMR has also been found close to
the present best fit solution. Nevertheless, the more extensive current search
reveals many other stable solutions between 3:1 and 9:1~MMRs equally good in
terms of $1\sigma$. Thus the results of our previous work are not wrong in
general but rather incomplete --- it appears that the search performed was not
exhaustive enough to find all acceptable solutions.  Still, some of the unstable
solutions can be modified in order to obtain stable fits, possibly with slightly
increased $\Chi$. For instance, the stability maps in Fig.~\ref{fig:fig3} seem
to rule out the 4:1~MMR in the system what may be related to specific orbital
phases. After an appropriate change of these phases [still keeping $\Chi$ at
acceptable level] we could find also stable solutions close to the 4:1~MMR. In
order to do this in a self-consistent manner, a GAMP-like method would be
necessary. However, as we have shown, $\Chi$ is very flat in the interesting
range of orbital parameters. Then, the extensive enough GAMP search would
require a very significant amount of CPU time, so we decided not to do it.

\begin{table}
\label{tab:tab1}
\caption{
Astrocentric, osculating Keplerian parameters of the edge-on, best-fit,
self-consistent configuration of 14~Her system at the epoch of the first
observation, $t_0=$~JD~2459464.5956. The mass of the parent star is
1.0~$M_{\sun}$. See the text and Fig.~\ref{fig:fig3} for the error estimates. 
The RV offsets $V_{\idm{N2004}}$, $V_{\idm{B2006}}$, and $V_{\idm{W2007}}$ label
the data sets published in \citep{Naef2004,Butler2006,Wittenmyer2007},
respectively; the offset of $V_{\idm{N2004}}$ is given with respect to the mean
of RV in \citep{Naef2004}. ${\cal M}$ denotes the mean anomaly at the epoch
$t_0$.
}
\centering
\begin{tabular}{lcc}
\hline
\hline
parameter \hspace{1em}  
& \ \ \ \ \ \ \ \ planet  {\bf b} \ \ \ \ \  & \ \ \ \ \ planet {\bf c} \ \ \ \ \  
\\
\hline
$m$ [m$_{\idm{J}}$]  
			&   4.975   &    7.679   
\\
$a$ [AU] 		&   2.864   &   9.037  
\\
$P$ [days]              &   1766    &   9886  
\\
$e$     		&   0.359   &    0.184  
\\
$\omega$ [deg] 		&  22.230   &  189.076 
\\
${\cal M}(T_0)$ [deg]
			&  330.421  &  81.976
\\
$V_{\idm{N2004}}$ [m s$^{-1}$] 	& \multicolumn{2}{c}{-52.049} 
\\
$V_{\idm{B2006}}$ [m s$^{-1}$] 	& \multicolumn{2}{c}{-55.290} 
\\
$V_{\idm{W2007}}$ [m s$^{-1}$] 	& \multicolumn{2}{c}{-31.368} 
 \\
$\Chi$  		& \multicolumn{2}{c}{1.0824}
\\
rms~[m s$^{-1}$] 	& \multicolumn{2}{c}{8.15}
\\
\hline
\end{tabular}
\end{table}

Finally, using the algorithm~II, we try to resolve the topology of $\Chi$ with
respect to a varying mass of the star and inclination of the system  (both can
be free parameters in the model) but assuming that the system remains 
coplanar.  Adopting the uncertainty of the 14~Her mass $\sim 10\%$
\citep{Butler2006}, we choose $0.9\mbox{M}_{\sun}, 1.0\mbox{M}_{\sun},
1.1\mbox{M}_{\sun}$, and inclinations of orbital plane as $90^{\circ}$ (the
edge-on system), $60^{\circ}, 45^{\circ}, 30^{\circ}$.  Apparently, the
distributions of the best-fits are similar for all ($\mbox{M}_{\star},i$)-pairs.
Their quality in terms of $\Chi$ is comparable.

\begin{figure*}
   \centerline{
   \vbox{
   \hbox{
       \hbox{\includegraphics[width=2.56in]{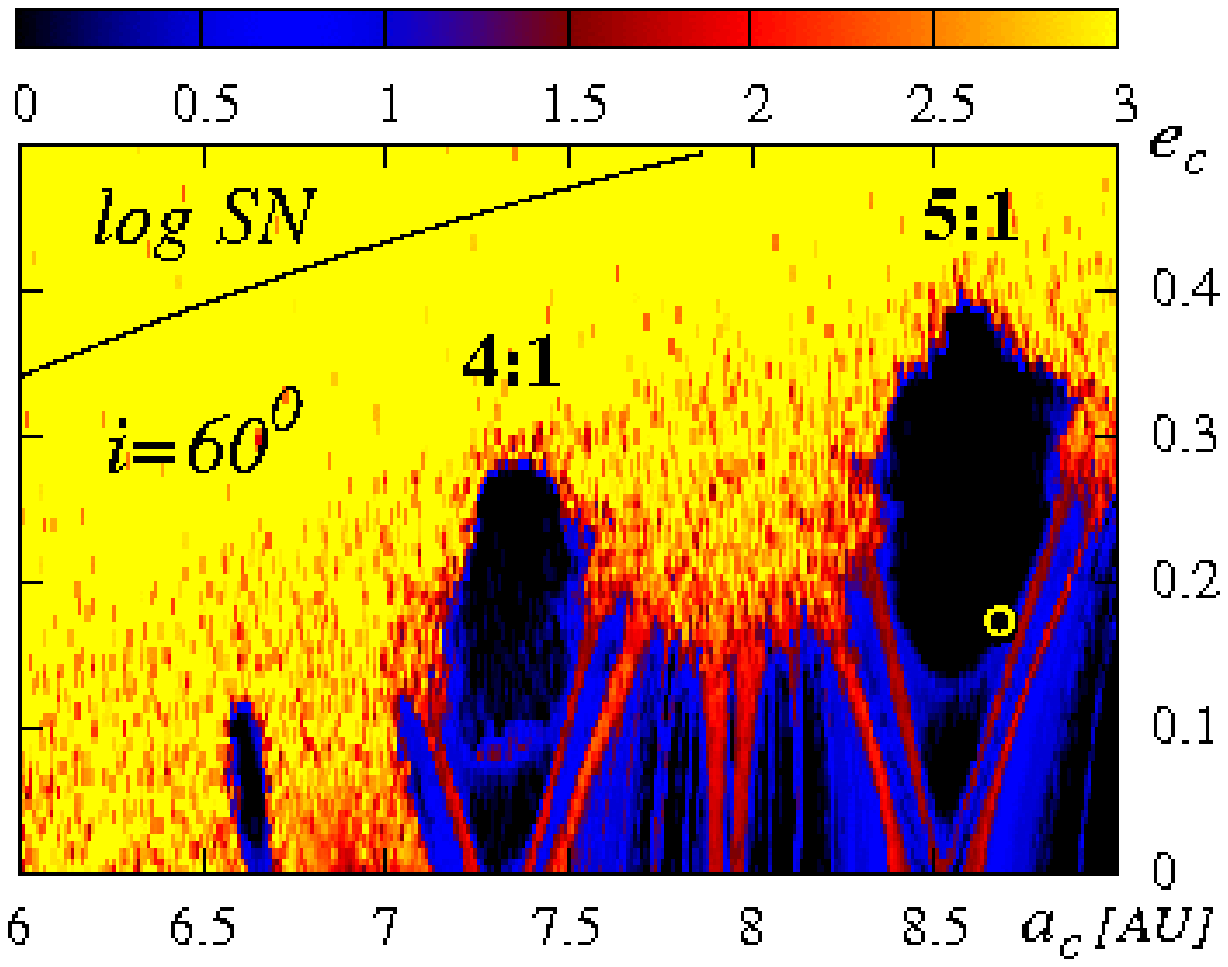}}
       \hspace*{1cm}
       \hbox{\includegraphics[width=2.56in]{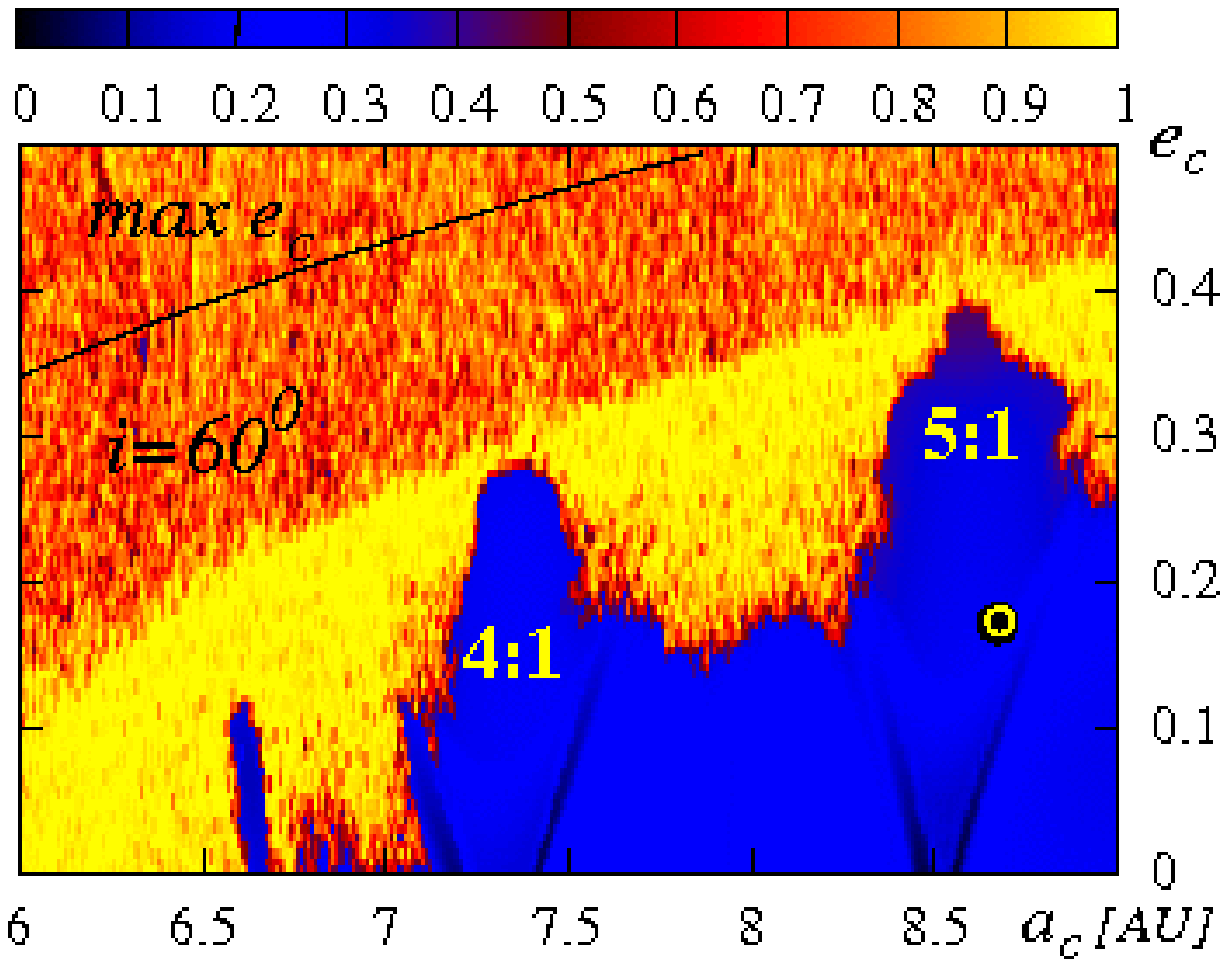}}
   }
   \hbox{
       \hbox{\includegraphics[width=2.56in]{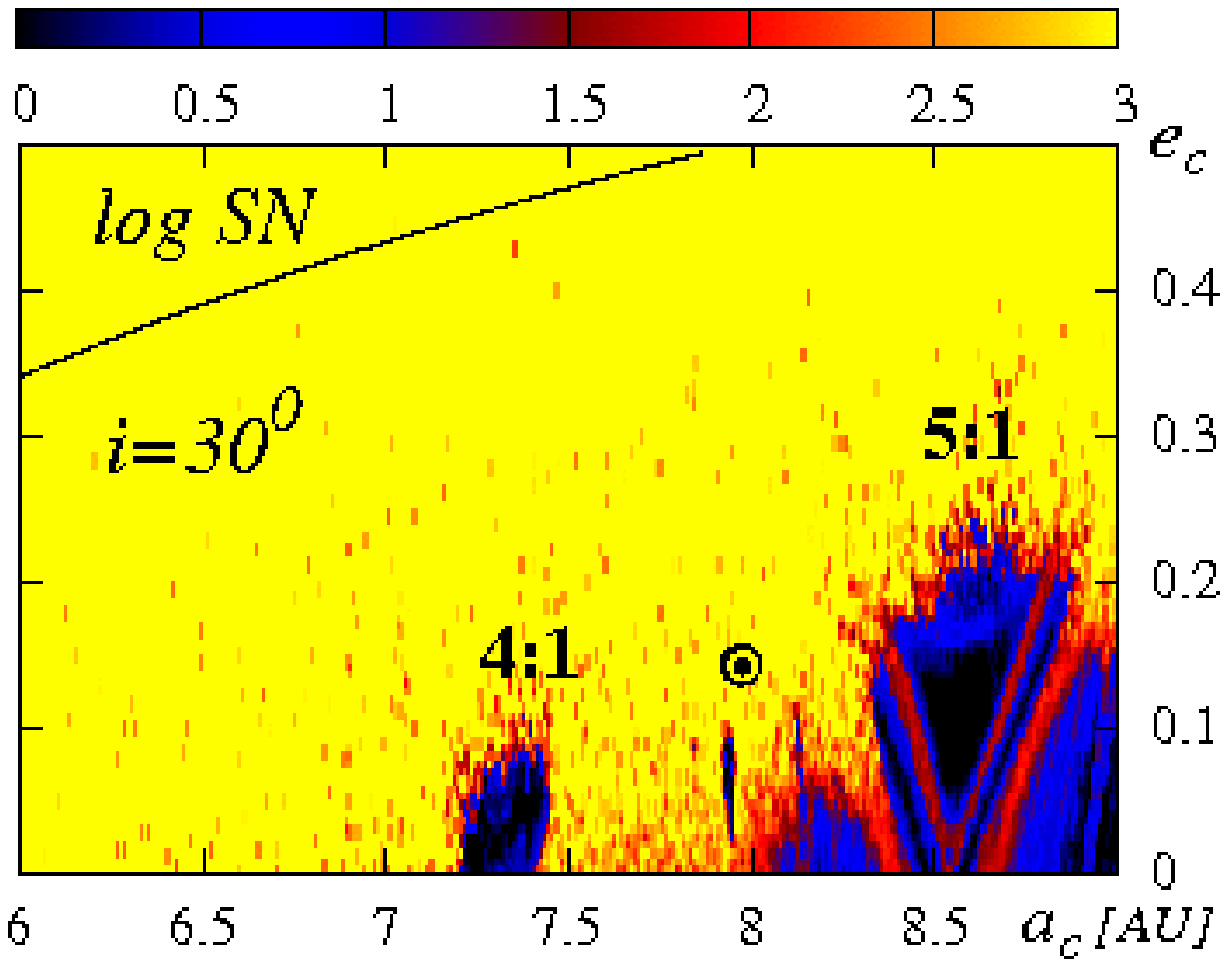}}
       \hspace*{1cm}
       \hbox{\includegraphics[width=2.56in]{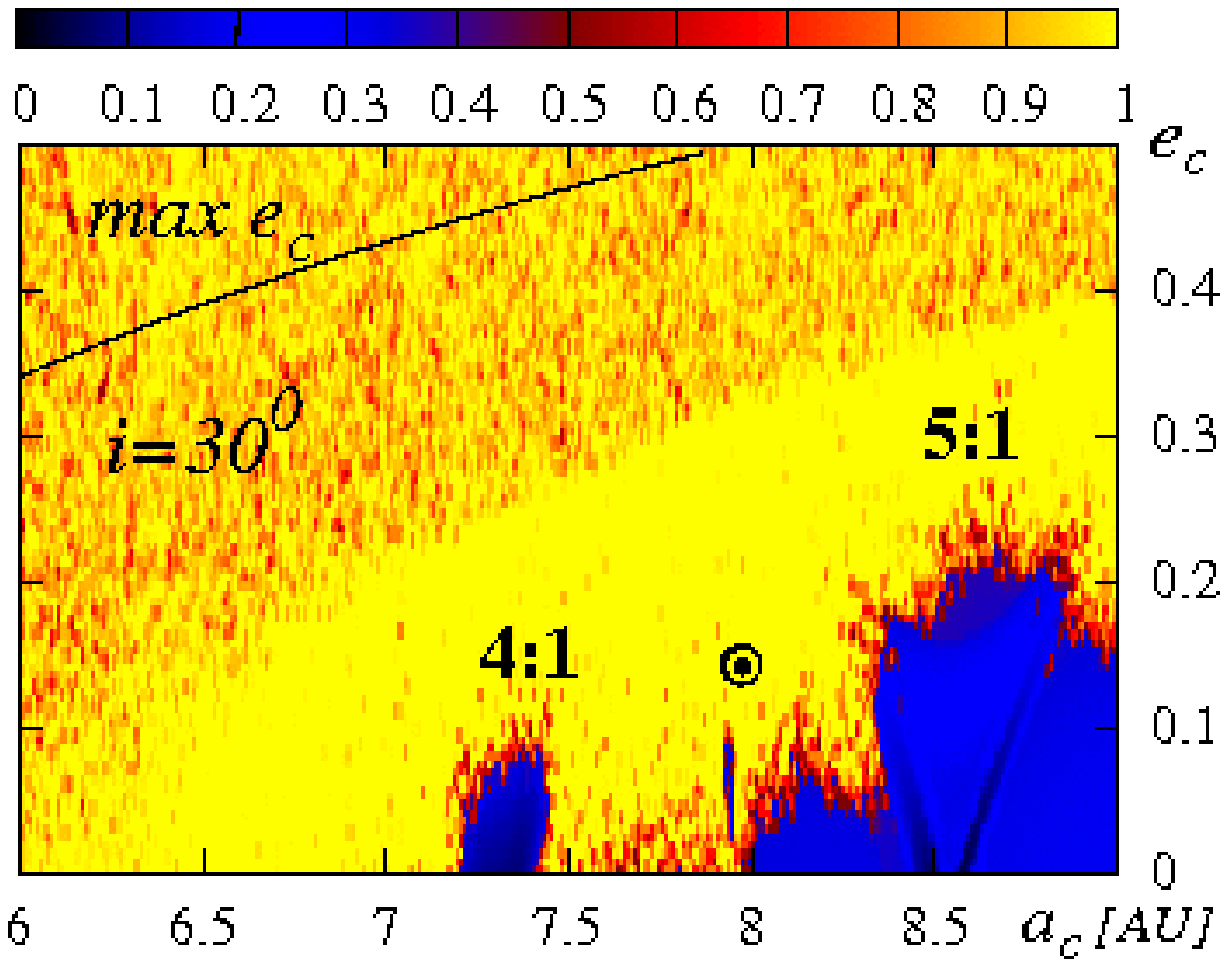}}
   }
   }
   }
   \caption{
     The dynamical maps in the $(a_{\idm{c}},e_{\idm{c}})$-plane in terms of the
Spectral Number ($\log SN$, the left panels), and $\max e_{\idm{c}}$ (the right
panels)  for the two-planet coplanar Newtonian models of the \stara{} system and
different inclinations of the orbital plane.  Colors used in the $\log SN$ map
classify the orbits --- black indicates quasi-periodic regular configurations
while yellow strongly chaotic ones. The maps are computed for the following
elements of the Jovian planets in terms of
tuples                                                   $(m~[m_{\idm{J}}],
a~[\mbox{AU}], e, \omega~[\mbox{deg}], {\cal M}(t_0)~[\mbox{deg}])$. The top row
is for $i=60^{\circ}$ and the osculating elements (5.747,    2.864,    0.359,  
22.271,  330.378) for the inner planet,  (7.343,    8.675,    0.171,  190.065,  
75.540) for the outer planet and   the velocity offsets (-43.458, -46.700,
-22.780)~[m/s]. The fit has $\Chi\sim 1.082$ and rms $\sim 8.15$~m/s. The bottom
row is for $i=30^{\circ}$ and the elements  (9.978, 2.868, 0.357, 22.498,
330.135) of the inner planet,  (8.581, 7.974, 0.142, 194.823, 58.154) of the
outer planet and the velocity offsets (-29.237, -32.485, -8.585)~m/s, $\Chi\sim
1.080$ and rms $\sim 8.13$~m/s.  The large crossed circles mark the parameters
of the best fits.  The low-order MMRs of the planets b and c are labeled. The
$SN$  integrations are conducted over $\sim 10^4 P_{\idm{c}}$. The resolution is
$500\times120$ data points. Compare with Fig.~\ref{fig:fig4}  for the edge-on
system.
   }
\label{fig:fig6}%
\end{figure*}

The significant differences between the best-fit configurations are related to
their stability.  To address this issue, we calculated the dynamical maps for
the best fits derived for the nominal mass of the parent star
($1~\mbox{M}_{\sun}$) and inclinations $90^{\circ}$ (the edge-on system,
Table~1), $60^{\circ}$, and $30^{\circ}$ (see
Fig.~\ref{fig:fig4},\ref{fig:fig6}). The $\log SN$ maps are accompanied by the
maximal eccentricity maps, $\max e_{\idm{c}}$, i.e., the eccentricity attained
by the orbit during the total integration time. The $\max e$ indicator makes it
possible to resolve a link between formally chaotic character of orbits with
their physical, short-term instability. In the regions where chaos is strong,
the maximal eccentricity grows quickly to~1, implying ejections or collisions
between the planets or with the star.

The nominal system lies in a stable zone between 5:1 and 6:1~MMRs (see
Fig.~\ref{fig:fig4}). However, for the inclination of $\sim 60^{\circ}$, the
best fit configuration appears locked in the 5:1~MMR. Overall,  the phase space
does not change significantly compared to the edge-on system
(Fig.~\ref{fig:fig4}, also Fig.~\ref{fig:fig2}). For the inclination of
$30^{\circ}$, the stable zone shrinks due to strong interactions caused by large
masses of the planets $\sim 10\mbox{m}_{\idm{J}}$ (see caption to
Fig.~\ref{fig:fig6}).   Moreover, the masses {\em are not} simply scaled by
factor $1/\sin i$, as one would expect, assuming the kinematic model of the RV.
Instead, already for $i\sim 30^{\circ}$, the mass hierarchy is reversed -- the
inner planet becomes more massive than the outer one. It is yet another argument
against the  kinematic model of the RV data. The best-fits derived for small
inclinations are confined to a strongly chaotic zone.  Comparing the dynamical
map for $i\sim 30^{\deg}$ with Fig.~\ref{fig:fig4}, we conclude that is unlikely
to find stable systems close to the formal best fit with $a_{\idm{c}} \sim
9$~AU  unless they are trapped in 5:1 MMR. The resonance island can be seen at
the edge of the bottom panels in Fig.~\ref{fig:fig6}. The planetary masses close
to the minimum of $\Chi$ remain smaller from the brown-dwarf limit even for
relatively small inclinations of the orbital plane.

\begin{figure*}
   \centerline{
   \hbox{
       \hbox{\includegraphics[width=3.3in]{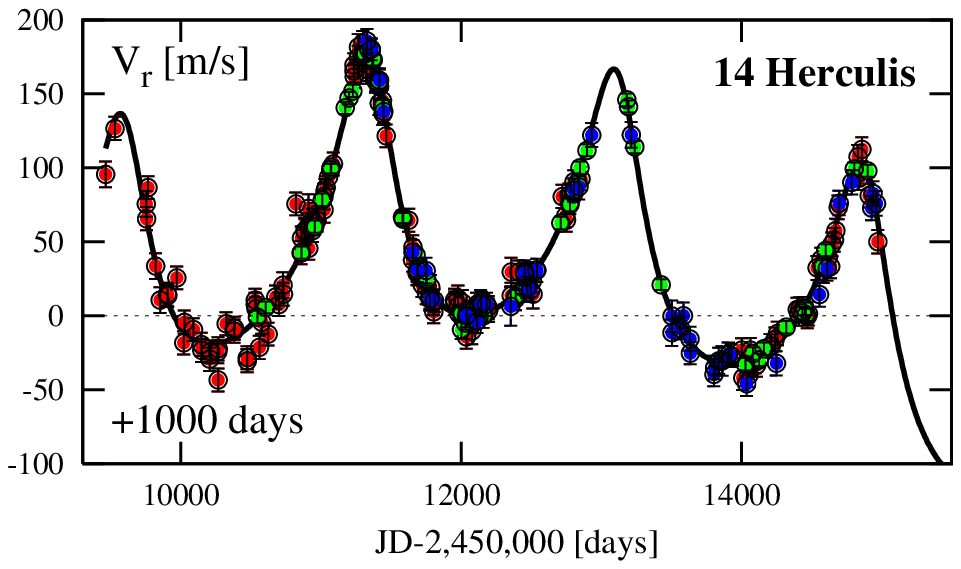}}
       \hspace*{1cm}
       \hbox{\includegraphics[width=3.3in]{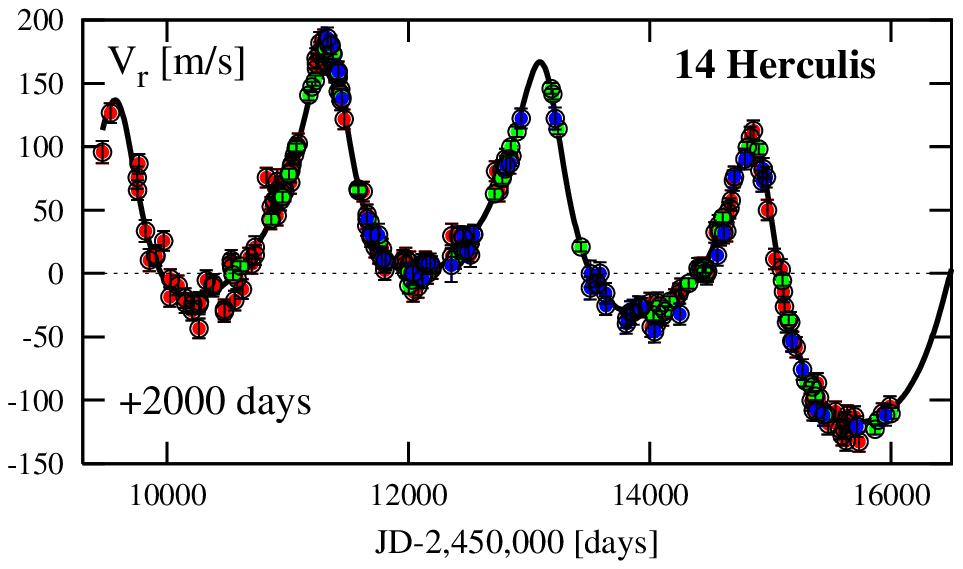}}
   }
   }
   \caption{
 Synthetic RV data and the RV curve of the corresponding best fit to the
two-planet model.  The left panel is for the time series extended over 1000~days
after the last real observation, the right panel is for the synthesized data
over 2000~days after the last real data point. Sampling frequency and errors of
the artificial observations are similar to those of the real data. 
}
\label{fig:fig7}
\end{figure*}

\begin{figure*}
   \centerline{
   \hbox{
       \hbox{\includegraphics[width=2.56in]{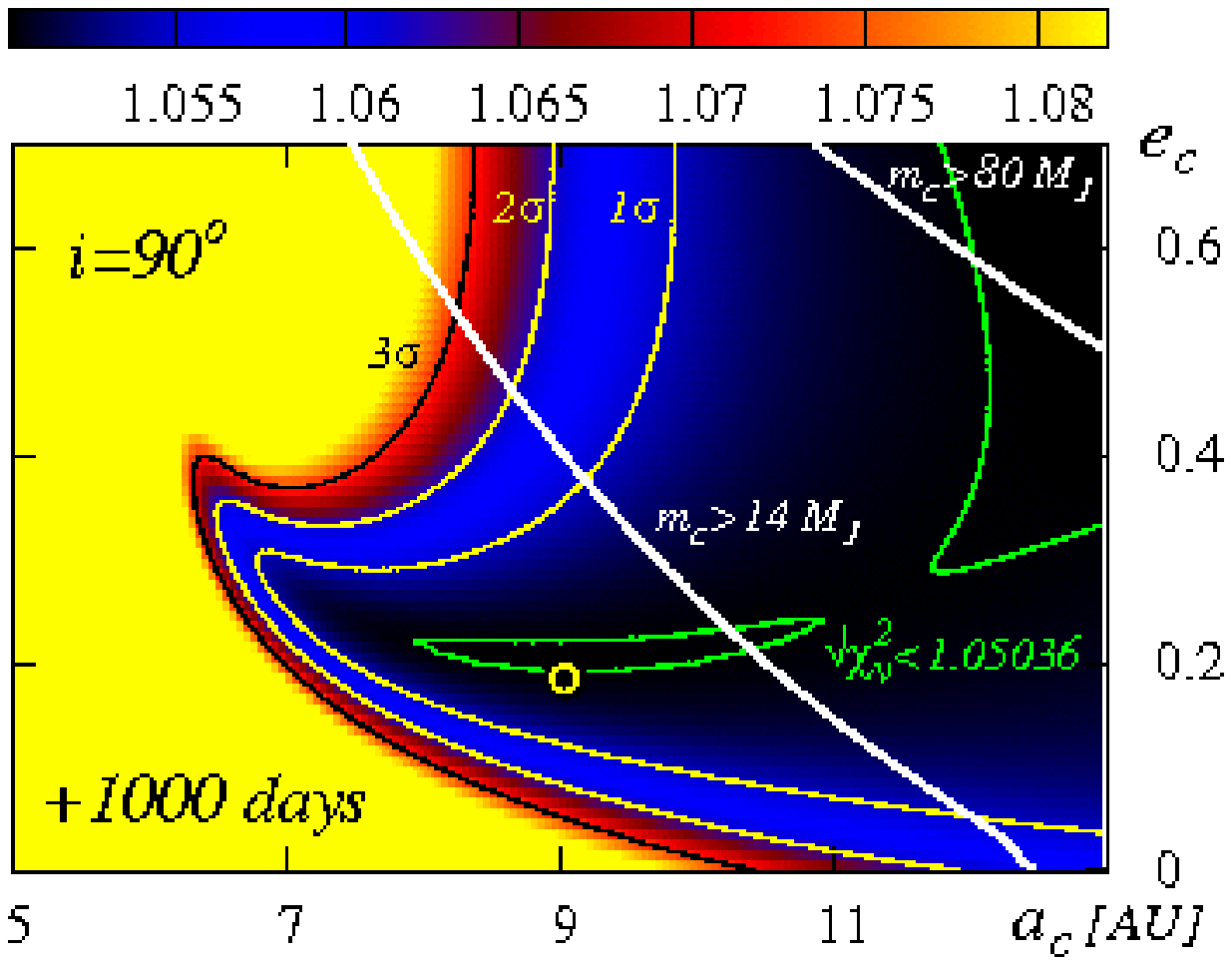}}
       \hspace*{1cm}
       \hbox{\includegraphics[width=2.56in]{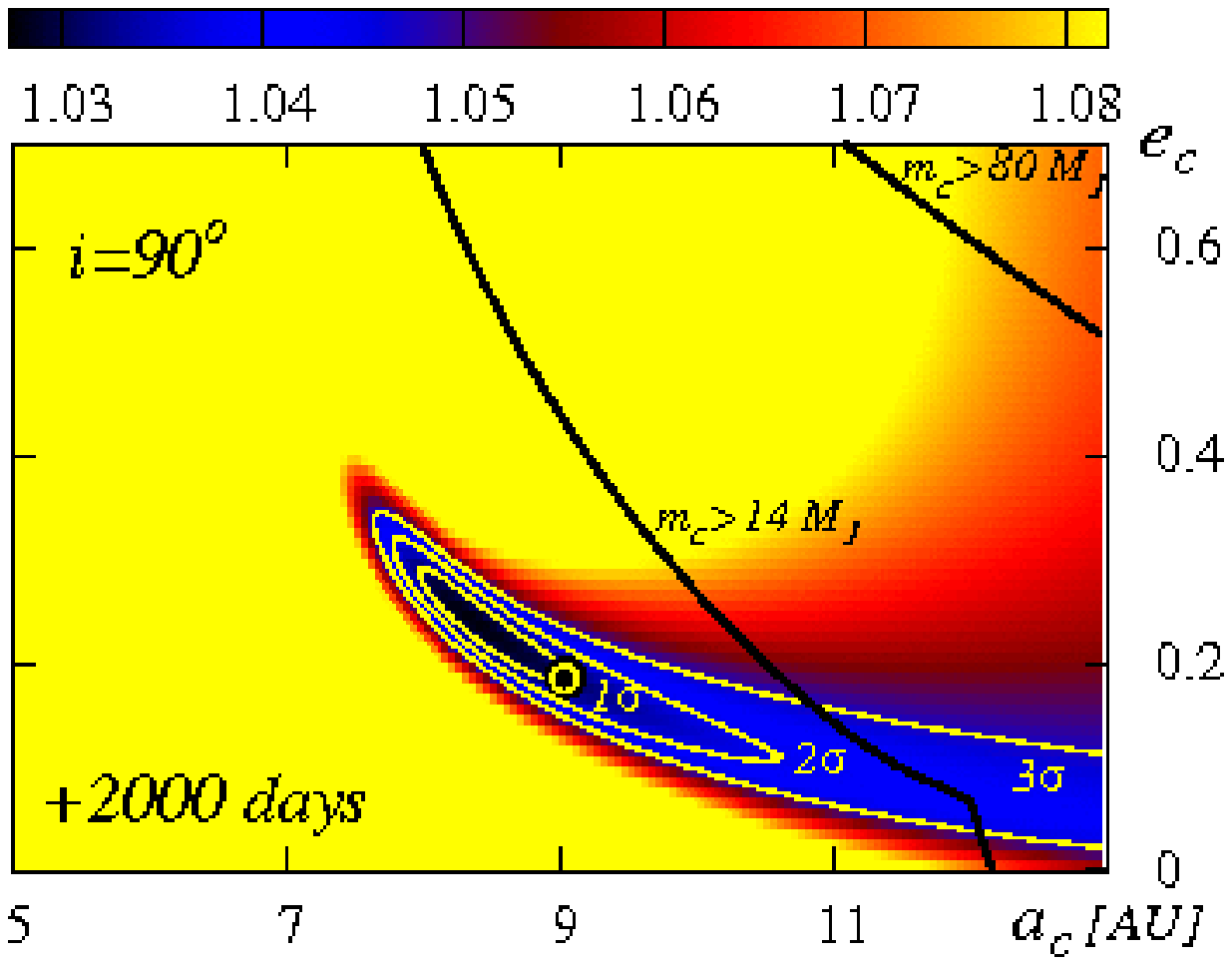}}
   }
   }
   \caption{
   The topology of $\Chi$ in the $(a_c,e_c)$-plane of the best fits to the
synthetic data sets shown in Fig.~\ref{fig:fig7}. The left panel is for
artificial observations spanning additional 1000~days after the date of the last
real data point, the right panel is for the additional 2000~days period.
Contours mark $1\sigma,2\sigma$ and $3\sigma$-levels of $\Chi$. Thick lines
(white in the left panel, black in the right panel) are for the mass limits of
$14~\mbox{m}_{\idm{J}}$ and $80~\mbox{m}_{\idm{J}}$, respectively. Circle marks
the position of the nominal best-fit  in Table~1 (the ``real'' system).
}
\label{fig:fig8}
\end{figure*}

Still, the parameters of the outer planet are not well constrained and it is
clear that many years of new observations are required to fix the elements of
the putative outer companion. To simulate the future RV analysis of the system,
we prepared two synthetic sets of observations, choosing the elements of Table~1
as the parameters of the ``real'' system. We computed the synthetic RV for that
system and added Gaussian noise to these data with parameters as from the real
observations. Next, we selected the data points sampling them with a similar
frequency to that in the real  measurements. We analyzed two sets: one extended
over +1000~days and the second one extended over +2000~days after the last
moment in the real observations.  Next, for both synthetic data sets
(Fig.~\ref{fig:fig7}), we scanned the $\Chi$ space with the algorithm~II in
order to resolve the best-fit system, moreover, in this experiment, no stability
constraints were imposed. The results are shown in two panels in
Fig.~\ref{fig:fig8}. Curiously, after the additional $\sim 3$~yr observations,
the parameters of the outer planet cannot be still determined without doubt.
After $\sim 6$~yr, the space of $\Chi$ shrinks significantly and we may expect
that after such a time, the semi-major axis of the outer companion can be fixed
with $1\sigma$ range of $\sim 1$~AU.

\section{Secular dynamics of the best-fit system}
%
Although the RV data permit for different dynamical states of the system, we can
analyse the secular evolution of the best fit coplanar and edge-on configuration
(Table~1) and the solutions in its neighborhood. Due to the significant
uncertainty of the orbital elements of outer planet, such analysis cannot
provide exhaustive conclusions but we can obtain some qualitative view of the
secular behaviour of the system. It can be easily extended when more RV data
will be available. 

As we have demonstrated, the parameters of the inner planet as well as the
eccentricity and orbital  phase of the putative outer companion can be
relatively well constrained. The most unconstrained parameter in the edge-on, 
coplanar system is $a_{\idm{c}}$. Yet we should remember that the observations 
constrain very roughly the inclinations; and we did not include the 
longitudes of nodes as free parameters in the model, keeping the planets  in
coplanar orbits. Still, we are in relatively a good situation because the 
careful analysis of the RV data makes it possible to reduce the dimension of the
phase space. The dynamical map in Fig.~\ref{fig:fig4} covering the shallow
minimum of $\Chi$ and the neighborhood of the formal best fit reveals its 
proximity to the 11:2~MMR. Several authors argue that in similar cases, 
\citep{Libert2007,Libert2005,Gallardo2005} in the first approximation  the
near-resonance effects have a small or negligible influence on the secular
dynamics.  Hence, neglecting the effects of the MMRs, we can recover its basic
properties with the help of the recent quasi-analytical averaging method by
\cite{Michtchenko2004}. It generalizes the classic  Laplace-Lagrange (L-L)
theory \citep{Dermott1999} or a more recent work by \cite{Lee2003}.  It relies
on a semi-numerical averaging of the perturbing Hamiltonian and makes it
possible  to avoid any series expansions when recovering the secular variations
of the orbital elements.   Then one obtains a very accurate secular model which
is valid up to large eccentricities by far beyond the limits of L-L classical
theory. The details of this  powerful  approach can be found in
\citep{Michtchenko2004,Michtchenko2006}. 

To study the secular dynamics, we follow \cite{Michtchenko2004} and describe the
system in terms of the canonical Poincar\'e variables. First, we obtain the
time-average of the canonical elements derived from the numerical integration of
the full equations of motion with the initial condition in Table~1. The mean
canonical semi-major axes of the system are then $\mean{a_{\idm{b}}} \sim
2.868$~AU and $\mean{a_{\idm{c}}} \sim 8.891$~AU.  The eccentricities are 
$\mean{e_{\idm{b}}}\sim 0.356$ and  $\mean{e_{\idm{c}}}\sim 0.191$. The initial
state of the system is also characterized by the difference of arguments of
periastron, $\Delta \varpi = \varpi_{\idm{b}}-\varpi_{\idm{c}} \sim 195^{\circ}$
with $1\sigma$ uncertainty $\sim 27^{\circ}$ that is estimated from the
statistics  of solutions derived for {\em fixed} masses and semi-major
axes  (parameters of the secular model), at their respective values
in Table~1.

\begin{figure*}
   \centerline{
   \hbox{
       \vspace*{-30mm}
   \hbox{
       \hbox{\includegraphics[width=4.4in]{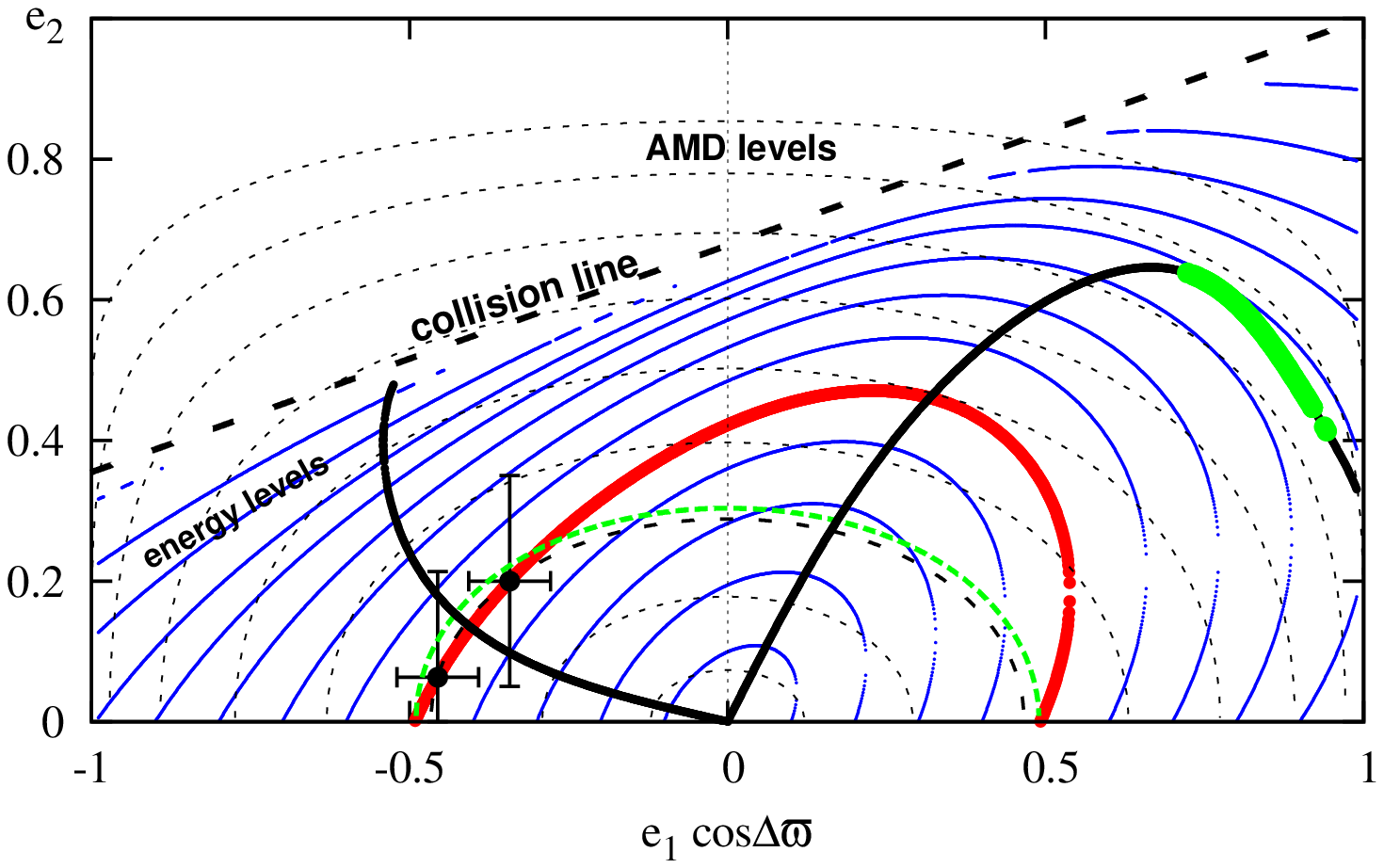}}
   }
   \hspace*{0mm}
   \vbox{
       \hbox{\includegraphics[width=2.0in]{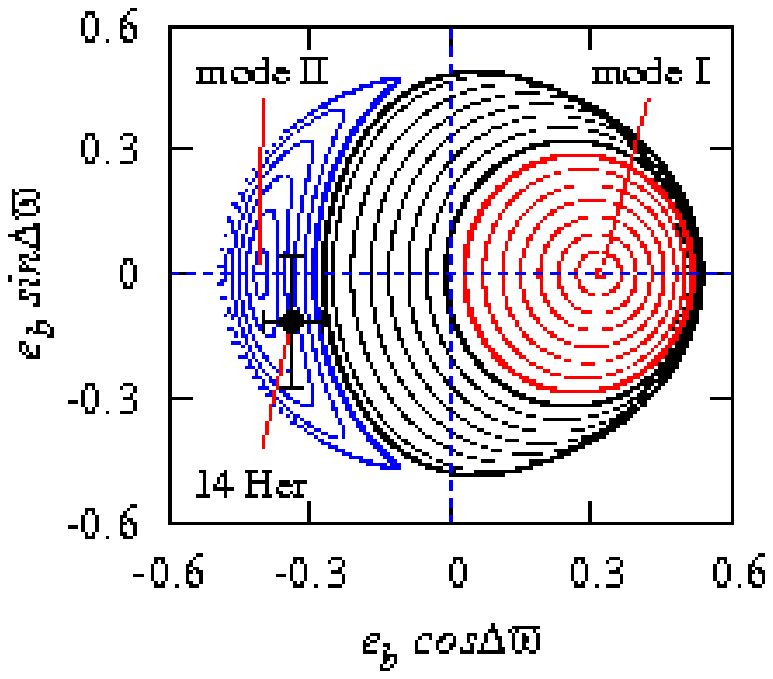}}
       \vspace*{-4mm}
       \hbox{\includegraphics[width=2.0in]{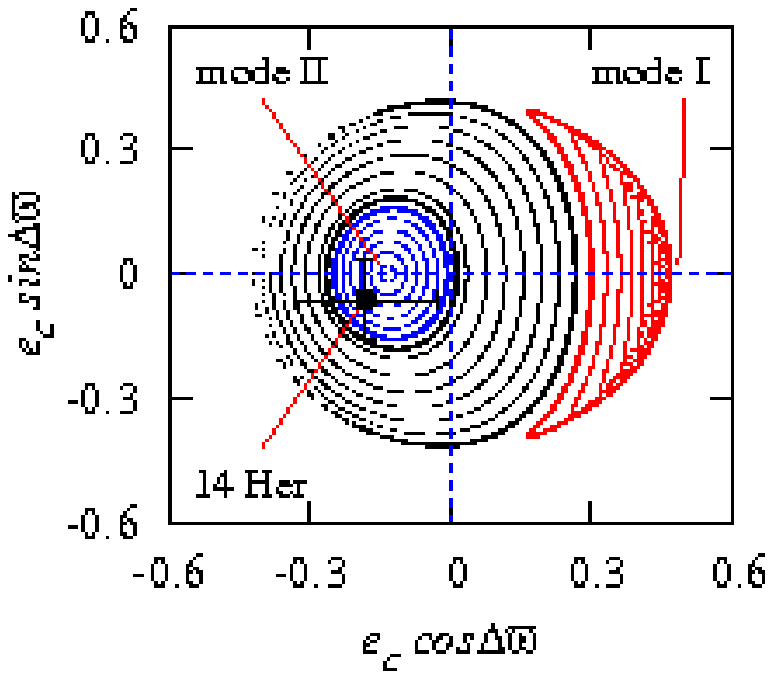}}
   }
   }
   }
   \caption{
  The secular evolution of the best-fit configuration (see Table~1). The
left
  panel is  for the energy levels of the averaged secular Hamiltonian
  $\mean{H}_{\idm{sec}}$ (blue, continuous curves) and the $AMD$ levels (black,
  dashed lines) on the representative plane of initial conditions.  The energy
  decreases  and $AMD$ increases  with increasing $e_{\idm{c}}$. The sign of
  $e_{\idm{b}}\cos\Delta\varpi$ is for initial conditions with
  $\Delta\varpi=0^{\circ}$ (+) or $\Delta\varpi=180^{\circ}$ (-), respectively.
  The level curves are computed for $m_{\idm{b}}/m_{\idm{c}}=0.648$ and
  $a_{\idm{b}}/a_{\idm{c}}=0.3226$.  The red thick curve marks the energy level
  of the nominal system, and filled circles indicate actual variation of the 
  eccentricities (through the angular momentum integral). The black thick lines
  mark the centers of librations of $\Delta\varpi$ around  $0^{\circ}$ 
  (mode~I) or $180^{\circ}$ (mode~II), respectively.  A green region (on the
  right, for positive $e_{\idm{b}}\cos\Delta\varpi$) is for the nonlinear
  secular resonance.  Panels in the left column are for the secular phase space of
  the planets~b and c at secular energy $\mean{H}_{\idm{sec}}$ of the nominal
  system in the  ($e_{\idm{b}}\cos \Delta\varpi, e_{\idm{b}}\sin
  \Delta\varpi)$-plane (the right-top panel) and   ($e_{\idm{c}}\cos \Delta\varpi,
  e_{\idm{c}}\sin \Delta\varpi)$-plane (the right-bottom panel), respectively. Two
  libration modes of $\Delta\varpi$ about the centers $\Delta\varpi = 0^{\circ}$
  (mode~I, red curves) and $\Delta\varpi = 180^{\circ}$ (mode II, blue curves)
  are labeled. Thick black lines are for the transition between the libration and
  circulation of $\Delta\varpi$. Initial mean parameters of the nominal systems
  are marked with filled circles. For a reference, we plot formal
  error bars derived from $1\sigma$ errors of the best fit parameters 
  in Table~1.
   }
\label{fig:fig9}%
\end{figure*}

Next, we compute the energy levels of the averaged Hamiltonian of the  system
$\mean{H_{\idm{sec}}}$ (Fig.~\ref{fig:fig9}, left panel)  in the characteristic
plane ($e_{\idm{b}}\cos\Delta\varpi,e_{\idm{c}}$) where $\Delta\varpi$ is set
either to $0^{\circ}$ or $180^{\circ}$. Let us note that $\Delta\varpi$ always
passes through these values during the orbital evolution related to two
distinct  libration modes of $\Delta\varpi$ in the coplanar system
\citep{Michtchenko2004}. The energy level of the nominal system is marked with
red line. The black thick lines mark the exact positions of libration centers 
(i.e., periodic solutions) of $\Delta\varpi$ around  $0^{\circ}$  (mode~I) and
$180^{\circ}$ (mode~II), respectively. A green part of the libration curve (on
the right half-plane of the energy diagram) is for the nonlinear secular
resonance of the system. The black dashed lines mark different levels of the
total angular momentum (expressed by the {\em Angular Momentum Deficit}, $AMD$).
The nominal system (marked with filled dots) is found in the region of the
anti-aligned apsides (close to the thin curve representing mode~II) with
$e_{\idm{c}}$ in the range $\sim (0.1,0.2)$. 

\begin{figure}
   \centerline{\hbox{\includegraphics[width=2.7in]{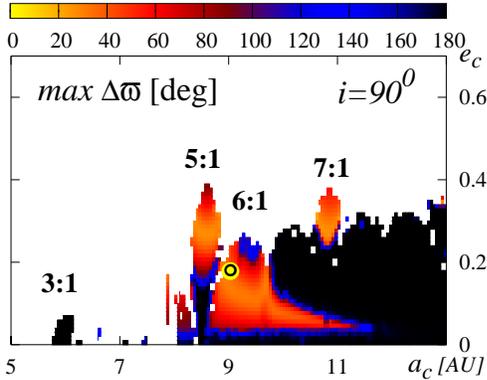}}}
   \caption{
The dynamical map in the $(a_{\idm{c}},e_{\idm{c}})$-plane in terms  of the 
$\max \Delta\varpi$ indicator for edge-on, coplanar system and assemble of fits
shown in Fig.~\ref{fig:fig3}, the top-left panel. The large crossed circles mark the
parameters of the  best fit solution in Table~1.  The low-order MMRs of the
planets b and c are labeled.  The integrations are conducted over  $\sim 10^4
P_{\idm{c}}$. 
}
\label{fig:fig10}
\end{figure}

To illustrate the secular dynamics in more detail, we computed the phase
diagrams of the system in the characteristic plane of $(e_{\idm{b}}
\cos\Delta\varpi,e_{\idm{b}}\sin\Delta\varpi)$ (the top-left panel in
Fig~\ref{fig:fig9}) and  in the $(e_{\idm{c}}
\cos\Delta\varpi,e_{\idm{c}}\sin\Delta\varpi)$-plane (the bottom-left panel in
Fig.~\ref{fig:fig9}). The red curves mark the oscillation around  libration
mode~I, blue curves are for  libration mode~II and black curves mark the
circulation of $\Delta\varpi$. The nominal system (marked with filled circles)
lies close to the center of mode~II with a relatively large amplitude of
$e_{\idm{b}} \sim 0.5$.  The librations of $\Delta\varpi$ are confined to small
semi-amplitudes $\sim 30^{\circ}$. The results of the secular theory are in an
excellent accord with the behavior of the full unaveraged system. Its dynamical
maps are shown in Fig.~\ref{fig:fig4}. The dynamical map of  $\max \Delta
\varpi$ (i.e., the maximal semi-amplitude of $\Delta \varpi$ attained during the
integration time) reveals a similar semi-amplitude of librations $\sim
30^{\circ}$ around the center of $180^{\circ}$ (and of course, the presence of
mode~II). This zone persists over wide ranges  of $(a_{\idm{c}},e_{\idm{c}})$.
In fact, almost the entire stable region extending at least up to 11~AU is
spanned by configurations involved in mode~II librations.   The error bars in
plots of Fig.~\ref{fig:fig9} are derived from formal estimates of $1\sigma$
errors of the parameters in Table~1 (we estimate the error of $\Delta\varpi \sim
27^{\circ}$; according with the analysis of stability, we adopted  the
uncertainty of $e_c$ as 0.15). The errors are significant but still, in the
statistical sense, the libration modes would mostly not change  within the error
ranges. This behaviour is confirmed in Fig.~\ref{fig:fig10} showing
$\Delta\varpi$ for each individual best-fit found   (see the top-left panel of
Fig.~\ref{fig:fig3}). Simultaneously, physically stable systems are close to
quasi-periodic configurations what follows from the comparison of the  $\max e$
and $\max \Delta\varpi$ indicators.   Both of them, as indicators of geometrical
or physical behaviour of the system in short-time scale, appear tightly
corelated to the $\log SN$ as the formal measure of the system stability. 

\section{Conclusions}
%

According to the results from the adaptive-optics imaging
\citep{Luhman2002,Patience2002},  there is no stellar-mass object in the 14~Her
neighborhood beyond $\sim 12$~AU.  This finding supports the hypothesis that the
RV trend may be attributed to a massive planet~c.  The orbital period of the
putative companion to the known Jovian planet~b cannot yet be  very well
constrained. Nevertheless the available data already reveal a very shallow
minimum od $\Chi$ in the ($a_{\idm{c}},e_{\idm{c}})$-plane of the initial
osculating elements. The minimum persists for reasonable combinations of the
parent star mass and the inclination of the system. Its dynamical character
strongly depends on that parameter influencing the planetary masses. Quite
surprisingly, for small  inclinations the mass hierarchy is reversed and the
inner planet becomes more massive than its distant companion. Also the positions
of the nearby MMRs are significantly affected. Depending on the inclination, the
system may be  locked in the low-order 9:2, 5:1 or 6:1~MMR. We can also conclude
that the kinematic model of  the RV is already not adequate for the
characterization of the system and further analysis of the RV would be done best
within the self-consistent Newtonian model.

The dynamical analysis enables us to derive some limits on the orbits elements
and the stability of the system. No stable configurations are found with a
period ratio smaller than $\sim 3$. This limit is shifted towards larger values
when the inclination grows. For the inclination of $30^{\circ}$, the best fit
solutions are localized in a strongly chaotic and unstable zone. This
constitutes a dynamically derived argument against a small inclination of the 
system. It is likely larger than $30^{\circ}$-$40^{\circ}$. Allowing for some 
speculation, the presence of the best-fits in a robustly stable zone supports 
the hypothesis of highly inclined two-planet configuration in the 14~Her
system.  Our attempts to fit one more planet to the RV data that could explain
the scatter  of the data points close to the minima of the RV curve
(Fig.~\ref{fig:fig5}), did not lead to a meaningful improvement of the fit. The
third planet would  have a weak RV signal at the level of a few m/s and a
short-period eccentric orbit. Finally, the analysis of all available RV data of
14~Her enables us to revise and extend some of the conclusions from
\cite{Gozdziewski2006a} --- an isolated and very well defined best-fit solution
cannot yet be  found to properly describe the orbital configuration of 14 Her. 

{\bf Acknowledgements.}
We thank Eric Ford for a critical review, suggestions and comments 
that greatly improved the manuscript. 
This work would not be possible without the access to the precision RV data
which the discovery teams make available to the community. This work is
supported  by the Polish Ministry of Science and Education through grant
1P03D-021-29. M.K. is also supported by the Polish Ministry of Science and
Education through grant N203 005 32/0449.

\bibliographystyle{mn2e}
\bibliography{ms}
\end{document}